\documentclass[graybox]{svmult}

\usepackage{mathptmx}       
\usepackage{helvet}         
\usepackage{courier}        
\usepackage{type1cm}        
\usepackage{makeidx}         
\usepackage{graphicx}        
\usepackage{multicol}        
\usepackage[bottom]{footmisc}
\usepackage{amsmath}	
\usepackage{amssymb}    
\usepackage{array}	 

\makeindex             

\def\prd{PRD\,}

\def\apj{ApJ\,}
\def\apjl{ApJ\,}
\def\aap{A\&A\,}
\def\mnras{MNRAS\,}
\def\apjs{ApJS\,}
\def\pasp{PASP\,}
\def\pasa{PASA\,}
\def\apss{Ap\&SS\,}
\def\aj{AJ\,}
\def\nat{Nature\,}
\def\araa{ARA\&A\,}
\def\procspie{Proc.\,SPIE\,}

\newcommand{\vt}{\textrm{v}}
\newcommand{\bt}{\beta}
\newcommand{\pr}{\partial}

\newcommand{\vsh}{\vt_s}
\newcommand{\xsh}{x_s}
\newcommand{\msh}{m_s}

\newcommand{\vBO}{\vt_{\rm bo}}
\newcommand{\rhoBO}{\rho_{\rm bo}}
\newcommand{\tBO}{t_{\rm bo}}
\newcommand{\EBO}{E_{\rm bo}}

\newcommand{\Rstar}{{R_*}}
\newcommand{\vmax}{\vt_{\max}}
\newcommand{\aBB}{a_{BB}}
\newcommand{\Av}{A_{\vt}}
\newcommand{\Linst}{L_{\rm inst}}
\newcommand{\Lobs}{L_{\rm obs}}
\newcommand{\Ztbar}{\langle Z^2\rangle}

\newcommand{\Abar}{\langle A\rangle}

\begin{document}

\title*{Shock breakout theory}
\author{Eli Waxman \& Boaz Katz}
\institute{
Eli Waxman \at Weizmann Institute of Science, Rehovot, Israel \email{eli.waxman@weizmann.ac.il}
\and Boaz Katz \at Weizmann Institute of Science, Rehovot, Israel \email{boaz.katz@weizmann.ac.il}
}
%
%
\maketitle

\abstract{
The earliest supernova (SN) emission is produced when the optical depth of the plasma lying ahead of the shock, which ejects the envelope, drops below $\approx c/\vt$, where $\vt$ is the shock velocity. This "breakout" may occur when the shock reaches the edge of the star, producing a bright X-ray/UV flash on time scales of seconds to a fraction of an hour, followed by UV/optical "cooling" emission from the expanding cooling envelope on a day time-scale. If the optical depth of circumstellar material (CSM) ejected from the progenitor star prior to the explosion is larger than $c/\vt$, the breakout will take place at larger radii, within the CSM, extending its duration to days time scale. The properties of the early, breakout and cooling, emission carry unique signatures of the structure of the progenitor star (e.g. its radius and surface composition) and of its mass-loss history. The recent progress of wide-field transient surveys enable SN detections on a day time scale, and are being used to set unique constraints on the progenitors of SNe of all types. This chapter includes:
\\ \noindent(i) A derivation of the properties of non-relativistic breakout bursts from H and He envelopes, and of
\\ \noindent(ii) the cooling envelope emission for H, He and C/O envelopes;
\\ \noindent(iii) A discussion of the constraints on progenitor properties that may be inferred from observations;
\\ \noindent(iv) A shorter discussion of CSM and relativistic breakouts focused on open theoretical issues;
\\ \noindent(v) A concise overview of what we have learned from observations so far, and of advances in observational capabilities that are required in order to make further significant progress.
}

\section{Introduction}
\label{sec:Intro}

The earliest emission of electro-magnetic radiation from a supernova (SN) explosion is associated with the "shock breakout" (for early work on this subject see\cite{Colgate68,Grassberg71,Colgate75,LasherChan75,Imshennik77,Falk78,KleinChevalier78,LasherChan79,Epstein81,EnsmanBurrows92}). As the radiation mediated shock (RMS), that drives the ejection of the SN envelope, expands outwards, the optical depth of the material lying ahead of it decreases. When the optical depth drops below $\approx c/\vt$, where $\vt$ is the shock velocity, radiation escapes and the shock dissolves. As long as the outer boundary of the ejecta does not expand significantly (by tens of percents), the radiation from deeper layers cannot escape and the emitted flux drops significantly. The breakout is expected to take place once the shock reaches the edge of the star, producing a bright X-ray/UV flash on time scales of seconds to a fraction of an hour, followed by UV/optical emission from the expanding cooling envelope on a day time-scale. Alternatively, if there is sufficient amount of circumstellar material (CSM) ejected from the progenitor star prior to the SN explosion, e.g. by a steady stellar "wind" or by an episodic ejection of an outer envelope shell, the breakout may take place at larger radii, within the CSM, provided that the CSM optical depth is larger than $c/\vt$ (for early work on this subject see \cite{Grassberg71,FalkArnett73,FalkArnett77,Chevalier82,ChevalierFransson94}). In this case, the breakout time scale may be extended to days. In CSM breakouts, the RMS is converted to a collisionless shock that expands further into the CSM, converts an increasing fraction of the kinetic energy of the ejecta to radiation and produces high energy photons and neutrinos.

During the next few days after breakout, as the ejecta expands, radiation from deeper layers escape with gradually declining temperature and slowly declining luminosity (envelope cooling emission). During the first few days, the radiation escapes from mass elements that were at the outer parts of the progenitor's envelope. The emission depends on the radius of the progenitor and on the expansion velocity, and is nearly independent of the structure and mass of the outer envelope. These early parts of the SN light curve, dominated by breakout and post-breakout cooling emission, precede the more widely observed and studied longer time scale SN emission powered by radioactive decay and shock energy deposited deep in the ejecta. The properties of the early emission carry unique signatures of the structure of the progenitor star (e.g. its radius and surface composition) and of its mass-loss history close to the explosion. Studying the early emission therefore provides unique information on the SN progenitors and their pre-explosion evolution (see \S~\ref{sec:observations}), which cannot be directly inferred from later time observations. This information is highly instructive for the study of the supernova explosion mechanisms,
which are not fully understood despite many years of research (for recent reviews see \cite{Smartt09CC-prog-rev,Janka12CC-exp-rev,Burrows13CC-exp-rev,Hillebrandt13Ia-rev,Maoz_Ia_14}).

Until recently, only a small number of SNe were detected early enough to enable such analysis of the early emission. However, the progress of wide-field optical transient surveys is changing this situation. Existing surveys (iPTF \cite{Rau_PTF09,Gal-Yam_iPTF11}, Pan-STARRS \cite{Pan-STARRS02}, ASAS-SN \cite{ASAS-SN14}) provide SN detections on day, or even shorter, time scales, and upcoming surveys (ZTF \cite{Law_ZPTF09}, LSST \cite{LSST09}) will provide higher quality data (earlier detections and wider spectral coverage) for a larger number of events (ZTF will provide roughly a dozen shock cooling detections per year at $t<1$~d at g-band starting 2017, and LSST will roughly double this rate \cite{Ganot14}), thus enabling a more systematic study. A wide field UV survey, as proposed e.g. by the ULTRASAT satellite (\texttt{http://space.gov.il/en/node/1129}, \cite{Sagiv14}), would significantly enhance the ability to constrain the properties of the progenitor and its environment, by providing early measurements at wavelengths which are near or below the spectral peak. Such UV observations are necessary, for example, for a robust and accurate determination of the progenitor's radius (see \S~\ref{sec:ObsTheory}).

Recent theoretical work \cite{Blinnikov98_93J,MM99,Blinnikov00_87A,Utrobin0799emBO,NS10,Piro10,DessartHillier10NumericBO,KasenNumeric11,RW11,Sapir11,Tominaga11,Katz12,SKWspec13,Tolstov13} provides a rather complete and self-consistent theoretical description of the emission of radiation during and following non-relativistic breakouts from stellar surfaces, with accurate analytic approximations for progenitors with polytropic outer envelopes \cite{RW11,Sapir11,Katz12,SKWspec13}. Both the breakout and post-breakout cooling emission are not sensitive to the details of the density profile (the value of the polytropic index), and are determined by the stellar radius $R_*$, the ratio of ejecta energy to its mass, $E/M$, and the opacity $\kappa$ (which depends on the envelope's composition). Early SN observations were used to set, utilizing the theoretical analyses, important constraints on the progenitors of SNe of type Ia, Ib/c and II, as discussed in some detail in \S~\ref{ssec:surface_obs}.

While the high energy breakout burst is expected to precede the optical emission in every supernova, an unambiguous identification of a non-relativistic breakout burst from the surface of a supernovae progenitor has not been achieved yet. One of the challenges is the fact that for most supernovae, the peak of the emitted spectrum is expected to be at photon energies of tens of eV (see section \S\ref{sec:PlanarT}) which are highly absorbed by the ISM. For small progenitors such as blue super giants, the peak frequency may exceed 1~keV (with a lower flux) and may thus be observed with little ISM absorption. A few past and existing X-ray telescopes have sufficient sensitivity, field of view and accumulated run time to allow the detection of few to tens of breakouts \cite{SKWspec13,CalzavaraMatzner04}, including the Roentgensatellit (ROSAT), Chandra X-ray Observatory (CXO) and the High Throughput X-ray Spectroscopy Mission (XMM-Newton). While few attempts to identify the $\sim100$ seconds time scale X-ray bursts in the archival data of ROSAT have been made \cite{Vikhlinin98,Greiner00}, with the aim of detecting Gamma-ray burst afterglows, a systematic search for supernova breakout bursts in existing archival data is yet to be preformed. In fact, as explained in \S~\ref{ssec:surface_obs}, a few to tens of events may be detectable in the existing data of XMM-Newton and such a search is warranted and may lead to the exciting discovery and characterization of a population of breakouts.

Recent observations lead to increasing interest in CSM breakouts (see \S~\ref{ssec:CSM}, \S~\ref{ssec:CSM_obs}),
which are considered as possible explanations of (at least part of) the new class of "super-luminous" SNe, of "double peak" SNe,
of low-luminosity gamma-ray bursts and X-ray flashes associated with SNe, and of the early part of the emission of SNe of type IIn. The canonical picture of the pre-explosion evolution of the massive progenitors is challenged by the inferred large mass loss episodes closely preceding the stellar explosion (see \S~\ref{ssec:CSM}). A complete quantitative derivation of the spectra of radiation produced in such breakouts is, however, not yet available (see \S~\ref{ssec:CSM}). Analytic analyses are challenged, for example, by the inherent non-steady nature of the shock structure, which evolves on a dynamical time scale. Current numeric calculations do not describe several processes that significantly affect the radiation field, including inelastic Compton scattering, the separation of electron and proton temperatures at high shock velocity, and the generation of high energy particles and photons following the formation of a collisionless shock. This limits the ability to test the CSM breakout explanation of the above mentioned phenomena, to discriminate between models, and to derive quantitative constraints on the progenitors and on their environment.

This chapter is organised as follows. In section \S\ref{sec:HydrodynamicProfiles} the assumed profiles of the progenitors at the outer parts of the envelope and the approximated hydrodynamic evolution of the shock prior to breakout are described. In section \S\ref{sec:Planar} the properties of non-relativistic breakout bursts from H and He envelopes are derived. In section \S\ref{sec:spherical}, the properties of the cooling envelope emission are derived for H, He and C/O envelopes. In \S~\ref{ssec:ObsTheBO} and in \S~\ref{ssec:ObsTheCool} we summarize the theoretical results derived in \S\ref{sec:Planar} and in \S\ref{sec:spherical} respectively, explaining how they can be used to describe the properties of the breakout and post-breakout emission, and how they may be used to derive constraints on progenitor properties from observations. Open issues in the theory of CSM and relativistic breakouts are discussed in \S~\ref{ssec:CSM} and \S~\ref{ssec:relativistic} respectively. In \S~\ref{sec:observations} we provide a brief overview of what we have learned from observations so far, and discuss advances in observational capabilities that are required in order to make further significant progress.

\section{Pre-breakout hydrodynamic profiles}\label{sec:HydrodynamicProfiles}

In this chapter we consider the emission on time scales of seconds to days. At these times the emission is dominated by the outer shells of the ejecta, which carry a small fraction of the ejecta mass $M$ \cite{WaxmanCampana07},
\begin{equation}\label{eq:deltam}
\delta_m\equiv\delta M/M<10^{-2},
\end{equation}
and which were located initially (prior to the explosion) near the stellar surface, at $r_0$ typically satisfying
\begin{equation}\label{eq:delta}
\delta\equiv(R_*-r_0)/R_*\ll1,
\end{equation}
where $R_*$ is the radius of the progenitor. At these early times, the properties of the escaping radiation are nearly independent of the detailed structure of the pre-explosion envelope, and are completely determined by $R_*$, by the opacity $\kappa$ and by the typical ejecta velocity
\begin{equation}
\vt_*=\sqrt{E/M},
\end{equation}
where $E$ is the energy deposited in the ejecta.

We assume that the pre-explosion density profile, $\rho_0(r_0)$, is well approximated at the outer layers of the progenitor star, where the mass $\delta M$ lying between $r_0$ and $R_*$ is negligible with respect to the stellar mass, by \cite{Chandrasekhar39}
\begin{equation}
    \label{eq:init_dnsty_prfl}
    \rho_0(r_0) = f_\rho \overline{\rho}_0 (1-\delta)^{-n}\delta^n,
\end{equation}
with $n=3$ for radiative envelopes (assuming uniform opacity and molecular weight) and $n=3/2$ for efficiently convective envelopes (assuming a polytropic equation of state, $p\propto\rho^{5/3}$). Here, $\overline{\rho}_0\equiv M/(4\pi/3)R_*^3$ is the pre-explosion average ejecta density, and $f_\rho$ is a numerical factor of order unity that depends on the detailed envelope structure \cite{Calzavara04}. We show below that the results depend only very weakly on the value of $f_{\rho}$ and $n$ and are therefore likely insensitive to deviations from the specific profile assumed, eq. \eqref{eq:init_dnsty_prfl}. Under this approximation, the fraction of the ejecta mass lying above $r_0$ is
\begin{equation}
\label{eq:dmAsdl}
    \delta_m(\delta) \equiv M^{-1}\int^{R_*}_{(1-\delta)R_*} {\rm d}r 4 \pi r^2 \rho_0(r)= \frac{3 f_{\rho}}{n+1}
    \left[1+O(\delta)\right]\delta^{n+1}.
\end{equation}
Since $\delta_m/\delta\propto\delta^n\ll1$ for small $\delta$, the approximation of eq.~(\ref{eq:init_dnsty_prfl}) holds up to significant values of $\delta$ (as long as the assumptions of uniform opacity and molecular weight / polytropic equation of state hold).

The velocity of the SN shock within the envelope is well approximated \cite{MM99} by an interpolation between the spherical Sedov--von Neumann--Taylor self similar solution \cite{VonNeumann47,Sedov59,Taylor50} and the planar Gandel'Man-Frank-Kamenetskii--Sakurai self similar solutions \cite{GandelMan56,Sakurai60},
\begin{equation}
    \label{eq:vs_nonrel0}
    \vsh(r_0) = \Av \left[ \frac{E}{M(r_0)} \right]^{1/2} \left[ \frac{M(r_0)}{\rho r_0^3} \right]^{\beta_1},
\end{equation}
where $M(r_0)$ is the ejecta mass enclosed within $r_0$ ($M(R_*)=M$), $\Av = 0.8$ and $\beta_1 = 0.2$. For $\delta_m\ll1$ we have
\begin{equation}
    \label{eq:vs_nonrel}
    \vsh(\delta) = \Av \vt_*
    \left( \frac{4\pi}{3f_{\rho}} \right)^{\beta_1} (1-\delta)^{(n-3)\beta_1}\delta^{-\beta_1 n}.
\end{equation}
eq.~(\ref{eq:vs_nonrel}) provides an approximate description of the dependence of $\vsh$ on $\delta$ at large $\delta$ values, and an accurate description for $\delta\ll1$. It also provides an approximate determination of the velocity normalization of the asymptotic profile, $\vsh\propto\delta^{-\beta_1 n}$, in terms of $\vt_*$. In what follows, we derive the emitted luminosity and spectrum for times at which the emission is dominated by shells with $\delta\ll1$, using only the leading order $\delta$ terms in eqs.~(\ref{eq:dmAsdl}) and~(\ref{eq:vs_nonrel}).

\section{Breakout burst}
\label{sec:Planar}
\subsection{Introduction}
\label{sec:intro}
As long as the shock propagates deep in the interior of the progenitor star, radiation cannot escape. Once the shock approaches the surface of the star, radiation starts to leak out and at the same time the outer layers start to expand outwards. As long as the distance that the outer-most shell has moved out is much smaller than $\Rstar$, the expansion has planar geometry and the optical depth of each mass shell is approximately constant. In this phase, radiation can only escape from a thin outer shell with an optical depth of order $c/\vBO$, where $\vBO$ is the breakout velocity of the shock as it approaches the surface (see \S\ref{sec:BreakoutBol} for a precise definition of $\vBO$). The first light from a supernovae is therefore a burst of the radiation deposited by the shock in the outermost shell \cite{Colgate68,Colgate75, LasherChan75, KleinChevalier78, LasherChan79, EnsmanBurrows92, MM99} with mass
\begin{equation}\label{eq:breakoutmass}
\delta M\sim \frac{c}{\vBO} \kappa^{-1} \Rstar^2\sim 5\times 10^{-5}\vt_{\rm bo, 9}^{-1}R_{13}^{2}\kappa_{0.34}^{-1} M_{\odot},
\end{equation}
releasing an energy of order
\begin{equation}\label{eq:PlanarErough}
\EBO\sim \delta M\vBO^2\sim 10^{46}\vt_{\rm bo, 9}R_{13}^{2}\kappa_{0.34}^{-1} \rm erg.
\end{equation}
Here, $\kappa=0.34\kappa_{0.34}{\rm cm^2/g}$ is the opacity, $\vt_{\rm bo}=10^9\vt_{\rm bo, 9}~\rm cm/s$, and $\Rstar=10^{13}R_{13}~\rm cm$. In this section the precise calculation of the properties of this burst are described and the main results are provided.

For Red and Blue supergiants, the density in the outer shell is of order $\rho\sim 10^{-9}-10^{-8}\rm g/cm^3$ and the breakout velocity is of order $\vBO\sim10^9-10^{10}\rm cm/s$ [see eqs.\eqref{eq:MVParam_v}, \eqref{eq:PlanarMVParam_rho}] implying that temperatures exceeding $50$~eV are obtained [eqs. \eqref{eq:PlanarTVnumeric},\eqref{eq:PlanarVTanalytic}]. In such conditions the gas, assumed to be mainly hydrogen and helium, is fully ionized and the scattering-dominated opacity is independent of temperature, $\kappa=0.40(1-0.5Y_{\rm He}){\rm cm^2/g}$ where $Y_{\rm He}$ is the He mass fraction. This allows a significant simplification as the evolution of the bolometric light and the hydrodynamic profiles can be calculated independently of the temperature (e.g. \cite{Weaver76}). The evolution of the temperature profiles and the spectral shape of the emitted radiation can later be calculated by solving the radiation transfer problem using the previously calculated hydrodynamic profiles.

One difficulty with calculating the properties of the breakout burst is that there is still significant hydrodynamic evolution during the emission as the shells are accelerated by the radiation. This is in contrast with later times during the supernovae emission where the ejecta is freely costing to a very good approximation. Moreover, the region where the radiation is emitted from is of the same order as the shock transition layer. The radiation mediated shock cannot be treated as a discontinuity but its structure needs to be calculated in a self consistent way.

In section \S\ref{sec:PlanarBol}, the calculation of the bolometric properties of the breakout burst are described, while in \S\ref{sec:PlanarT} the calculation of the spectral properties are described. In each of these two sections the relevant equations of motion are provided. In each case the equations are first applied to the time independent structure of the radiation mediated shock as it propagates deep in the star and then to the time-dependent problem of the breakout burst itself.

\subsection{Bolometric and Hydrodynamic Properties of the Breakout Burst}\label{sec:PlanarBol}

The analysis of the hydrodynamic evolution and the bolometric radiation emitted in the burst is preformed under the following approximations, which are valid during the breakout in most supernovae explosions (e.g. \cite{LasherChan79}):
\begin{itemize}
\item The velocities are non relativistic, $\bt=\vt/c\ll 1$;
\item The internal energy (and pressure) of the matter are neglected;
\item Photon transport is described by diffusion with constant Thomson opacity of fully ionised gas $\kappa$;
\end{itemize}

\subsubsection{Equations determining the bolometric and hydrodynamic properties}\label{sec:PlanarEOMrhop}
It is useful to work with Lagrangian equations with the spatial coordinate chosen as the mass $m$ (per unit area) from the surface,
\begin{equation}\label{eq:m}
m=\int_\Rstar^r\rho dr=-\frac{\delta_mM}{\Rstar^2}.
\end{equation}
The coordinate $m$ is negative in the star, grows towards the outside and is zero at the surface. The optical depth to the surface is given by
\begin{equation}
\tau=-\kappa m
\end{equation}
and is sometimes used as the spatial coordinate instead.
The spatial position $x$ is accordingly chosen as
\begin{equation}
x=r-\Rstar=-\delta\times \Rstar.
\end{equation}
The equations of motion are given by (e.g. \cite{Lasher79,Sapir11})
\begin{align}\label{eq:PlanarLEOM}
\pr_t x&=\vt, \cr
\pr_{t}\vt&=-\pr_mp,\cr
\pr_t(e/\rho)&=-\pr_mj-p\pr_m\vt,\cr
j=&-\frac{c}{3\kappa}\pr_me,\cr
\end{align}
where $e$ is the thermal energy per unit volume, $p$ is the pressure and $j$ is the bolometric flux of radiation. The equation of state is
\begin{equation} \label{eq:EOS}
e=3p.
\end{equation}

\subsubsection{Radiation mediated shock: density and pressure}\label{sec:PlanarRMSrhop}
As long as the shock is far from the edge of the star, $\tau\gg c/\vsh$, its structure changes on timescales which are much longer than the shock crossing time.  The structure and conditions are well described by the steady state solution of a shock traversing an infinite, cold, homogenous medium with density $\rho_0$. Gas traversed by the shock is heated to an internal energy (per unit volume) $e_{\rm ps}$, accelerated to $\vt_{\rm ps}<\vsh$, and compressed to $\rho_{\rm ps}$ by radiation within the shock transition layer, where the subscript $_{ps}$ stands for post-shock.
The post-shock conditions can be readily found by mass, momentum and energy conservation (respectively),
\begin{equation}\label{eq:Conservation equations}
\rho_0\vsh=\rho_{\rm ps}(\vsh-\vt_{\rm ps}), ~~~p_{\rm ps}=\rho_0\vsh\vt_{\rm ps}, ~~~e_{\rm ps}=\rho_{\rm ps}\vt_{\rm ps}^2/2,
\end{equation}
where the last equation is evident in a frame moving with the post-shocked material. By dividing the last two equations, using the equation of state, $e_{\rm ps}=3p_{\rm ps}$, and substituting $\rho_0\vsh$ with $\rho_{\rm ps}(\vsh-\vt_{\rm ps})$ we obtain the post-shock velocity
\begin{equation}\label{eq:vps}
\vt_{\rm ps}=\frac67\vsh
\end{equation}
and post-shock density
\begin{equation}\label{eq:rhops}
\rho_{\rm ps}=7\rho_0.
\end{equation}
The kinetic energy per mass and thermal energy per mass are equal and given by
\begin{equation}\label{eq:eps}
\frac{e_{\rm ps}}{\rho_{\rm ps}}=\frac12\vt_{\rm ps}^2=\frac{18}{49}\vsh^2.
\end{equation}
Note that the equality of kinetic and thermal energy for strong shocks (cold upstream) is directly implied by the third equation among eqs. \eqref{eq:Conservation equations} and holds for any equation of state.

The transition in the hydrodynamic properties between the upstream conditions and the downstream conditions is smooth and occurs across a region with optical depth of order the diffusion length $\tau\sim c/\vsh$. The transition profile is obtained by solving Equations \eqref{eq:PlanarLEOM} for a stationary shock profile, where the hydrodynamic quantities $A=\vt,\rho,p,j$ depend only on the separation from the shock,
\begin{equation}\label{eq:StationaryShock}
A(m,t)=f_A(m-\msh(t))
\end{equation}
where $\msh(t)$ is the shock mass coordinate, which grows at a constant rate
\begin{equation}
\frac{d\msh}{dt}=\rho_0\vsh.
\end{equation}
Note that there is freedom in choosing the point in the profile which marks the position of the shock. In the expressions below, $\msh(t)$ is chosen as the mass coordinate where the velocity is half of the final post-shock value $\vt(m=\msh(t))=\vt_{\rm ps}/2=6\vsh/14$.
The resulting velocity, density and pressure profiles are given by (e.g. \cite{Lasher79,Blandford81} and references therein)
\begin{equation}\label{eq:RMS_v}
\vt=\frac{\vt_{\rm ps}}{1+e^{3\tilde m}},
\end{equation}
\begin{equation}\label{eq:RMS_rhop}
\rho=\frac{\rho_0\vsh}{\vsh-\vt},~~p=\rho_0\vsh\vt,
\end{equation}
where
\begin{equation}
\tilde m\equiv\kappa\vsh(m-\msh)/c.
\end{equation}

The spatial displacement of each element in the profile from the shock coordinate is given by:
\begin{equation}\label{eq:RMS_x}
x-\xsh=\int_{\msh}^m \frac{dm}{\rho}=\frac{2c}{7\kappa\rho_0\vsh}\left[\ln\left(\frac{1+e^{3\tilde m}}{2}\right)+\frac12\tilde m\right].
\end{equation}

\subsubsection{Breakout: Bolometric properties}\label{sec:BreakoutBol}
Once the shock approaches a distance from the surface of the star, which is comparable to its own width, equations \eqref{eq:Conservation equations}-\eqref{eq:RMS_x} fail to capture the hydrodynamic profiles. In order to calculate the evolution, equations \eqref{eq:PlanarLEOM} need to be solved numerically. The properties of the breakout burst depend on the density and shock velocity at breakout as well as on the radius of the progenitor star. Note that the mass coordinate where breakout occurs, which satisfies $\tau\sim c/\vsh$, is only vaguely defined given that the width of the shock transition layer is comparable to the distance to the surface. A useful precise definition for the position of the shock (and shock velocity) at breakout for a given progenitor and explosion is defined as those satisfying  exactly $\tau=c/\vsh$ in a pure hydrodynamic solution, where diffusion is not included  (eq. \eqref{eq:PlanarLEOM} with $j=0$). In such solutions, which are known analytically for power-law profiles or can easily be obtained numerically otherwise, the shock is a discontinuity and has a precise position and velocity at any given time.
Once defined in this way, the properties of the shock breakout are completely determined by the progenitor radius $\Rstar$, the breakout velocity $\vBO$ and the initial density profile $\rho_0(\tau)$. It is useful to define the breakout density, $\rhoBO$, as the density at the breakout point
\begin{equation}\label{eq:rhoBO}
\rhoBO=\rho_0(\tau=c/\vBO),
\end{equation}
and to express the initial density profile as
\begin{equation}\label{eq:trho}
\rho_0(\tau)=\rhoBO\tilde\rho(\tau\vBO/c),
\end{equation}
were $\tilde\rho$ is a dimensionless function that describes the shape of the profile and satisfies $\tilde\rho(1)=1$. It turns out that the properties of the breakout flash are insensitive to the shape of the profile $\tilde\rho$, and are thus mainly set by $\Rstar,\vBO$ and $\rhoBO$.

Solving for $\tau=c/\vsh$ in the assumed profiles, eqs. \eqref{eq:init_dnsty_prfl} and \eqref{eq:vs_nonrel}, the following relations are obtained for $n=3$ (appropriate for a blue supergiant (BSG)) and for $n=3/2$ (appropriate for a red supergiant (RSG)):
\begin{align}\label{eq:MVParam_v}
\vBO/\vt_*&= 13 M_{10}^{0.16}\vt_{*,8.5}^{0.16}R_{12}^{-0.32}\kappa_{0.34}^{0.16} f_{\rho}^{-0.05}~~(n=3)\cr
&= 4.5 M_{10}^{0.13}\vt_{*,8.5}^{0.13}R_{13}^{-0.26}\kappa_{0.34}^{0.13} f_{\rho}^{-0.09}~~(n=3/2),
\end{align}
\begin{align}\label{eq:PlanarMVParam_rho}
\rhoBO&= 8\times 10^{-9} M_{10}^{0.13}\vt_{*,8.5}^{-0.87}R_{12}^{-1.26}\kappa_{0.34}^{-0.87} f_{\rho}^{0.29} \rm gr~cm^{-3}~~(n=3)\cr
&= 2.2 \times 10^{-9} M_{10}^{0.32}\vt_{*,8.5}^{-0.68}R_{13}^{-1.64}\kappa_{0.34}^{-0.68} f_{\rho}^{0.45} \rm gr~ cm^{-3}~~(n=3/2),
\end{align}
where $M_{\rm ej}=10 M_{10} ~M_{\odot}$, $R=10^{12}R_{12}\rm ~cm=10^{13}R_{13}\rm ~cm$, and $\vt_*=3,000\vt_{*,8.5}\rm ~km~s^{-1}$.

The timescale over which the flash is emitted from the surface is of order the crossing time of the shock width,
\begin{equation}
\tBO=\frac{c}{\kappa \rhoBO \vBO^2}=90 \kappa_{0.34}^{-1}\rho_{-9}^{-1}\vt_{\rm bo,9}^{-2}\rm s
\end{equation}
with $\rho=10^{-9}\rho_{-9}{\rm g/cm^3}$, $\vt_{\rm bo}=10^9\vt_{\rm bo,9}{\rm cm/s}$, and is typically much shorter than the light crossing time of the star.

It is useful to express the instantaneous luminosity $\Linst(t)$,
\begin{equation}\label{eq:PlanarLNorm}
\Linst(t)=4\pi \Rstar^2 \rhoBO\vBO^3 \mathcal{\tilde L}\left(\frac{t}{\tBO}\right),
\end{equation}
where $t=0$ is chosen as the time at which $\Linst$ peaks. The dimensionless function $\mathcal{\tilde L}$ depends on the shape of the density profile only and was calculated in \cite{Sapir11} for profiles with power-law indexes in the range $n=0-10$. Tabulated values are provided in their appendix. The dependence on $n$ is weak. The observed luminosity is not equal to the instantaneous luminosity due to the smearing caused by the light travel time. Moreover, even slight deviations from spherical symmetry may result in different shock arrival times at different positions on the surface. There are two robust properties of the breakout that are not sensitive to small deviations from spherical symmetry and that do not require the light travel effects to be taken into account. The first is the total emitted energy during the planar phase and the second is the (relatively) late time emission $\Rstar/c\ll t\lesssim\Rstar/\vBO$.

The total energy is given by
\begin{equation}\label{eq:PlanarEinfty}
\EBO=\int_0^{\infty}\Linst(t)dt=2.0\times 4\pi \Rstar^2\frac{\vBO c}{\kappa}=2.2\times 10^{47}R_{13}^2\vt_{\rm bo,9}\kappa_{0.34}^{-1} \rm erg~s^{-1},
\end{equation}
where the difference with respect to eq. \eqref{eq:PlanarErough} is the pre-factor $2.0$ which was numerically found in \cite{Sapir11, Katz12} and is accurate to better than $10\%$ for $1<n<10$.
At times much greater than $\tBO$, the luminosity follows $L(t)\propto t^{-4/3}$ \cite{Piro10,NS10}, and is approximately given by \cite{Sapir11, Katz12}
\begin{align}\label{eq:PlanarELApprox}
&\Lobs(t)\approx \Linst(t)= L_{\infty}\left(\frac{t}{\tBO}\right)^{-4/3}=3.0\times 10^{42}R_{13}^2\kappa_{0.34}^{-4/3}\vt_{\rm bo,9}^{1/3}\rho_{\rm bo,-9}^{-1/3}t_{\rm hr}^{-4/3}\rm erg~ s^{-1},\cr
\end{align}
where
\begin{equation}\label{eq:PlanarLinfty}
L_{\infty}\approx0.33\times 4\pi \Rstar^2\rhoBO\vBO^3,
\end{equation}
$\rhoBO=10^{-9}\rho_{\rm bo,-9} \rm gr~cm^{-3}$ and $t=1t_{\rm hr}$~hr.
eqs. \eqref{eq:PlanarELApprox} describe the emitted flux to an accuracy of better than  $30\%$ in $L(t)$  for $1<n<10$ and $1\ll t/\tBO<100$. The weak dependence on the parameters $\rhoBO$ and $\vBO$ implies that, if detected, this power law tail can be used for an accurate determination of the stellar radius. We note that for a constant density profile, $n=0$ (which is not directly relevant here), the luminosity declines faster than suggested in Eq. \eqref{eq:PlanarELApprox} and is given by $L(t)\propto t^{-9/8}$ \cite{Sapir11}.

Finally, in the adopted non-relativistic approximation in planar geometry an exact relation exists between the velocity of the outermost mass element and the emitted luminosity \cite{Lasher79, Sapir11},
\begin{equation}\label{eq:Planarv_to_Egen}
\vt(t)=\frac{\kappa}{c}\int_{-\infty}^t\mathcal{L}(t')dt'=\frac{\kappa E(t)}{4\pi R^2c}.
\end{equation}
Equation \eqref{eq:Planarv_to_Egen} simply states that photons that hit a given particle transfer all their momentum to the particle on average. It holds for any elastic scattering which has forward/backward symmetry, regardless of whether the diffusion approximation is valid or not.
In particular, the asymptotic value of the velocity of the surface is
\begin{equation}\label{eq:PlanarAssymptoticVelocity}
\vmax=\frac{\kappa \EBO}{4\pi R^2c}= 2.0\vBO,
\end{equation}
were we used the numerical pre-factor from eq. \eqref{eq:PlanarEinfty}.

\subsection{Temperature and spectrum at breakout}\label{sec:PlanarT}
We next consider the spectrum of the emitted breakout burst. At sufficiently early times, when the shock is far from the surface, mass elements traversed by the shock reach thermal equilibrium, $p_{\rm ps}=\aBB T_{\rm ps}^4/3$ and the postshock temperature is approximately given by
\begin{equation}\label{eq:PlanarTeq}
T_{\rm eq}=\left (\frac{18}7\rho_0\vsh^2/\aBB\right)^{1/4}\approx 66 \rho_{-9}^{1/4}\vt_9^{1/2}\rm eV,
\end{equation}
where  $\aBB=\pi^2/15(\hbar c)^{-3}$ is the Stefan Boltzmann energy density
coefficient. Thermal equilibrium requires the presence of a photon density $n_{\gamma}\approx p/T=\aBB T_{\rm ps}^3/3$. If the timescale for production of such photons (mainly by Bremsstrahlung) is much longer than the shock crossing time, thermal equilibrium is not achieved in the vicinity of the shock \cite{Weaver76,Katz10,NS10}. In such conditions, which are obtained at high velocities, the immediate postshock region has a high pressure set by the density and velocity and a low number of photons, resulting in temperatures which may be significantly higher than the equilibrium temperature resulting in the emission of hard-X rays or gamma-rays. As the shock approaches the surface, the value of the temperature in each element depends on the pressure which is set by the hydrodynamics and by the number of photons, which depends on the generation and diffusion of the photons.

While thermal equilibrium is not necessarily achieved, the photons and the plasma exchange energy efficiently through Compton scatterings. Within a diffusion length from the shock in the postshock region, a photon has about $\tau^2$ scatterings where $\tau\sim c/(\vsh-\vt_{\rm ps})\sim 7c/\vsh$ and can be up-scattered by a factor of $e^{y}$ in energy, where
\begin{equation}
y=\frac{4T}{m_ec^2}\tau^2\sim 35 \frac{T}{100\rm eV}\vt_9^{-2}
\end{equation}
is the Compton y parameter. For the parameters considered here, $y$ is significantly larger than unity and the radiation approaches an approximate Wein spectrum with a temperature equal to the electron temperature. This implies that to a good approximation the radiation can be described by two parameters, the energy density and the temperature (single photon approximation, \cite{Weaver76}).

\subsubsection{Equations determining the evolution of the temperature and emitted spectrum}\label{sec:PlanarEOMT}
The equations that describe the diffusion and generation of photons are given by \cite{Weaver76,SKWspec13}
\begin{equation}\label{eq:ngammajgamma}
\partial_t({n_{\gamma}/\rho})=-\partial_m j_{\gamma}+Q_{\gamma}(\rho,T)/\rho,
\end{equation}
where
\begin{equation}\label{eq:jgamma}
j_{\gamma}=-\frac{c}{3\kappa}\partial_m n_{\gamma}
\end{equation}
is the photon flux and
\begin{equation}\label{eq:Qgamma}
Q_{\gamma}=\frac{\alpha_e}{m_p}\kappa\rho^2 c\sqrt{\frac{m_ec^2}{T}}\Lambda(\rho,T)f_{\rm abs}
\end{equation}
is the photon generation rate per unit volume. $\alpha_e\approx 1/137$ is the fine structure constant,
\begin{equation}
\Lambda=\frac{\Ztbar}{\Abar}E_1(\lambda)\times[0.62-0.24\ln(\lambda)+6.6\times 10^{-4}\ln^2(\lambda)]
\end{equation}
is an effective gaunt factor, $\Ztbar=\sum Y_iZ_i^2$, and $\Abar=\sum Y_i A_i$ where $Z_i$, $A_i$ and $Y_i$ are the atomic number, atomic mass and ion fraction of ion $i$ respectively. $E_1(\lambda)=\int_{\lambda}^{\infty}dx e^{-x}/x\approx -0.5772-\ln(\lambda)+\lambda$ (accurate for small $\lambda$) and
\begin{equation}
\lambda=\frac{h\nu_c}{T}= 1.9\times 10^{-3}\sqrt{\frac{\Ztbar}{\Abar}}\rho_{-9}^{1/2}\left(\frac{T}{\rm keV}\right)^{-9/4},
\end{equation}
where $h\nu_c$ is the cutoff photon energy above which photons can up-scatter significantly before being absorbed (free-free absorption). Finally,
\begin{equation}
f_{\rm abs}=1-\frac{e}{\aBB T^4}
\end{equation}
is an approximate correction to account for free-free absorption that ensures that the photon density is constant in thermal equilibrium. Note that the composition enters only through the combination $\Ztbar/\Abar$, which for any mixture of Hydrogen and Helium equals unity $\Ztbar/\Abar=1$. The results below are obtained for this value.

The temperature is related to the pressure (known from the hydrodynamic solution of eqs. \eqref{eq:PlanarLEOM}) and the photon density by:
\begin{equation}\label{eq:PlanarTpnga}
T=\frac{p}{n_{\gamma}}.
\end{equation}
Note that in thermal equilibrium there is a $10\%$ error in equation \eqref{eq:PlanarTpnga} due to the photon degeneracy. This correction is ignored here.

\subsubsection{Radiation mediated shocks: Temperature}
\label{sec:PlanarRMST}
Equations \eqref{eq:ngammajgamma} and \eqref{eq:jgamma} can be analytically solved for the stationary hydrodynamic shock structure \eqref{eq:StationaryShock}-\eqref{eq:RMS_x} \cite{Katz10},
\begin{equation}\label{eq:PlanarRMSngamma}
n_{\gamma}(x)=\int_{-\infty}^{\infty}dx'Q_{\gamma}(x')G(x',x),
\end{equation}
where
\begin{equation}
G(x',x)=3\kappa\rho_0\frac{\vsh}{c}\int_{-\infty}^{\min(x,x')}\frac{e^{3\kappa\rho_0\vsh(x''-x)/c}}{\vsh-\vt(x'')}dx''.
\end{equation}
A self consistent solution for the temperature profile of a radiation mediated shock is obtained by preforming iterations on the Temperature profile \cite{Weaver76,Katz10}. Within each iteration, the photon generation rate $Q_{\gamma}$ is calculated throughout the profile using \eqref{eq:Qgamma}, then the photon density is calculated using \eqref{eq:PlanarRMSngamma} and finally a more accurate temperature profile is calculated using \eqref{eq:PlanarTpnga}.

An approximate expression for the maximal temperature, which is achieved in the vicinity of the hydrodynamic transition region, can be obtained by adopting the following simplifying approximations: 1. A constant post-shock velocity $\vt=\vt_{\rm ps}$, density $\rho=\rho_{\rm ps}$ and temperature $T=T_{\rm peak}$ (for production of photons), and therefore constant $Q_{\gamma}$; 2. A negligible contribution of photons from the pre-shocked region; 3. The number of photons is far from equilibrium so that $f_{\rm abs}=1$.  Under these conditions, equation \eqref{eq:PlanarRMSngamma} at the shock transition reduces to $n_{\gamma}=7Q_{\gamma}c/(3\kappa\vsh^2\rho_0)$, and using eqs. \eqref{eq:Conservation equations}-\eqref{eq:rhops}, \eqref{eq:PlanarTpnga} and eq. \eqref{eq:Qgamma} the following relation is obtained \cite{Katz10,NS10}:

\begin{align}\label{eq:PlanarVTanalytic}
\vsh&=\frac{7}{\sqrt{3}}\left(\frac{\alpha_em_e\Lambda(7\rho_0,T_{\rm peak})}{2m_p}\right)^{1/4}\left(\frac{T_{\rm peak}}{m_ec^2}\right)^{1/8}c\cr
&=3.7\times 10^9 \left(\frac{\Lambda(7\rho_0,T_{\rm peak})}{10}\right)^{1/4}\left(\frac{T_{\rm peak}}{\rm keV}\right)^{1/8}\rm cm/s.
\end{align}

For shock velocities $\vsh>2\times10^9 ~\rm cm/s$ and densities $\rho>10^{-9}{\rm g/cm^3}$ this approximation agrees with non-relativistic numerical calculations to about $20\%$. The same approximations can be used to estimate where deviations from thermal equilibrium are obtained. By equating eq. \eqref{eq:PlanarTeq} and eq. \eqref{eq:PlanarVTanalytic}, it is found that deviations from thermal equilibrium are expected for velocities exceeding \cite{Weaver76,Katz10,NS10}
\begin{equation}
\vsh\gtrsim1.5\times 10^{9}\Lambda^{4/15} \rho_{-9}^{1/30}\rm cm~ s^{-1}.
\end{equation}

\subsubsection{Breakout burst: Spectrum}
Equations \eqref{eq:ngammajgamma}-\eqref{eq:PlanarTpnga} were solved in \cite{SKWspec13} for power law profiles with indexes $n=3/2$ and $n=3$ applicable to Red and Blue supergiants. The evolution of the temperature with time are tabulated.  Here we focus on the most robust observational aspect, which is the integrated spectrum throughout the burst.
The energy emitted per logarithmic frequency,
\begin{equation}
\nu E_{\nu}=\int dt \nu L_{\nu}(t),
\end{equation}
peaks at a frequency, which is insensitive to the power-law index $n$ (at least for the calculated cases of $n=3$ and $n=3/2$) and is fitted by the following expression within the breakout velocity range $5\times 10^8\rm{cm/s}<\vBO<10^{10} \rm{cm/s}$ and density range of $10^{-11}\rm{g/cm^{3}}\rhoBO<10^{-7}\rm{g/cm^{3}}$ to an accuracy better than about $20\%$ \cite{SKWspec13},
\begin{equation}\label{eq:PlanarTVnumeric}
\log_{10}\left(\frac{h\nu_{\rm peak}}{\rm eV}\right)=1.4+\vt_{\rm bo,9}^{1/2}+(0.25-0.05\vt_{\rm bo,9}^{1/2})\log_{10}(\rho_{\rm bo,-9}).
\end{equation}
For velocities $3\times 10^9\rm{cm/s}<\vBO<10^{10} \rm{cm/s}$, this is equivalent to the analytic estimate of the post shock temperature eq. \eqref{eq:PlanarVTanalytic} to an accuracy of about $20\%$ in velocity if we assume $h\nu_{\rm peak}=3T$ and substitute the breakout velocity and density for the shock velocity and pre-shock density respectively.

The peak amount of energy per logarithmic frequency is about
\begin{equation}\label{eq:PlanarnuEnupeak}
\nu E_{\nu,\rm peak}\approx 0.9\times4\pi \Rstar^2 \frac{\vBO c}{\kappa}\approx 0.5 \EBO.
\end{equation}

These results are supported to by a more detailed calculation in which the single photon approximation is relaxed \cite{SapirHalbertal14} and the spectrum is calculated by solving the Kompaneets equation. Deviations by a factor reaching $1.5$ in the peak frequency of the integrated flux are obtained for low densities $\rho\lesssim 10^{-11}\rm{g/cm^{3}}$ and high velocities $\vBO\gtrsim 6\times 10^9\rm{cm/s}$. For the rest of the calculated range of velocities $10^9\rm{cm/s}<\vBO<6\times 10^{9} \rm{cm/s}$ and density $\rho=10^{-9}\rm{g/cm^{3}},10^{-7}\rm{g/cm^{3}}$ the deviations from the results of the single photon approximation are smaller than about $20\%$ in the peak frequency.

\section{Post breakout cooling envelope emission: The spherical phase}
\label{sec:spherical}
In this section we consider the emission from the cooling expanding shocked shell on a time scale of hours to days. At these times the shell has typically expanded to radii, which are significantly larger than the initial stellar radius $R_*$, and the emission is dominated by the outer shells of the ejecta, which carry a small fraction of the ejecta mass $M$, $\delta_m\equiv\delta M/M<10^{-2}$ \cite{WaxmanCampana07}, (see eq. \eqref{eq:deltam}), and which were located initially (prior to the explosion) near the stellar surface, at $r_0$ typically satisfying $\delta\equiv(R_*-r_0)/R_*<0.1$. At these early times, the properties of the escaping radiation are nearly independent of the detailed structure of the pre-explosion envelope, and are completely determined by $R_*$, by the opacity $\kappa$ and by the ratio $E/M$, where $E$ is the energy deposited in the ejecta.

The derivation given below is based on hydrodynamic ejecta profiles derived from the Gandel'Man-Frank-Kamenetskii--Sakurai self similar solutions \cite{GandelMan56,Sakurai60}, which provide an accurate description of the dynamics for $\delta\ll1$, and on the analysis of Matzner \& McKee \cite{MM99}, which provides an approximate determination of the normalization of the self-similar density and velocity profiles at $\delta\ll1$ in terms of the total ejecta mass and energy, $M$ and $E$ (see section \S\ref{sec:HydrodynamicProfiles}). As the ejecta continues to expand and the photosphere penetrates deeper to larger $\delta$ values, deviations from the self-similar description become significant. The results presented below are therefore accurate for
\begin{equation}\label{eq:t_valid}
  1.2\frac{(M/M_\odot)^{0.4}}{E_{51}^{0.5}}R_{*,13}^{1.3}\,{\rm\,hr}
                      <t<1.2\kappa_{0.34}^{0.5}\frac{(M/M_\odot)}{E_{51}^{0.5}} \,{\rm\,d}.
\end{equation}
Here, $E=10^{51}E_{51}$~erg, $R_*=10^{13}R_{*.13}$~cm and $\kappa=0.34\kappa_{0.34}{\rm cm^2/g}$. The lower limit is set by requiring the shell radius to exceed $3R_*$ (see eq.~(\ref{eq:non_rel_r_ph})), and the upper limit is set by requiring $\delta_m<10^{-2.5}$ (see eq.~(\ref{eq:ph_prop_nonrel})). For larger values of $\delta_m$, $\delta$ exceeds $\simeq0.1$ and the evolution of the expanding ejecta and of the escaping radiation depends on the detailed structure of the progenitor star (and is no longer accurately described by the self-similar solution).

Throughout this chapter we assume that the post-shock energy density is dominated by radiation. For very compact progenitors,
\begin{equation}\label{eq:IaR}
  R_*<10^{9.5} (M/1.4M_\odot)^{4/3}E_{51}^{-1}\,{\rm cm}
\end{equation}
as expected for SNe of type Ia, this assumption does not hold away from the immediate vicinity of the stellar edge \cite{RLW12Ia}. When the photosphere reaches regions in which the energy density is not dominated by radiation, the post-shock cooling emission is strongly suppressed (compared to the results derived in this chapter). This suppression is expected to occur at \cite{RLW12Ia}
\begin{equation}\label{eq:tIa}
  t\approx 1 E_{51}^{0.7}(M/1.4M_\odot)^{-0.6}(R/10^{8.5}{\rm cm})\,{\rm hr}.
\end{equation}

We first derive in \S~\ref{ssec:hydro} the density, velocity and pressure profiles of the (post-breakout) expanding stellar envelope. We then derive in \S~\ref{ssec:const_opacity} the luminosity and effective temperature of the escaping radiation for a constant (spatially and temporally independent) opacity, which is a good approximation for H dominated envelopes, since at the characteristic high temperatures and low densities H is nearly fully ionized and the opacity if dominated by Thomson scattering off free electrons. The model is extended in \S~\ref{ssec:recombination} to include the variation of the opacity due to recombination, which is important for He, C and O dominated envelopes. An approximate estimate of the ratio between color and effective temperatures is given in \S~\ref{ssec:Tc}. In \S~\ref{ssec:reddening} we explain how the effects of reddening may be determined, and hence corrected for, using multi-wavelength observations. Finally, we explain in \S~\ref{sec:ObsTheory} how early multi-wavelength observations of the spherical shock cooling phase may be used to determine $R_*$, $E/M$ and the reddening curve, as well as to constrain the composition of the outer envelope.

The derivations given below follow the analytic formalism of \cite{RW11}, that has been tested against numerical simulations and self-similar solutions and describes available observations well, and which includes a treatment of the effects of opacity variations due to recombination. Other analytical models (e.g. \cite{Chevalier92,NS10}) are limited to constant opacity, and provide broadly similar results (see \cite{Ganot14}). In particular, after appropriate corrections, the results of \cite{NS10} are in general agreement with those of \cite{RW11} (see \cite{Ganot14}).

\subsection{Hydrodynamic profiles}
\label{ssec:hydro}
As the radiation mediated shock passes through a fluid element lying at $r_0$, it increases its pressure to
\begin{equation}\label{eq:p0}
    p_0 = \frac{6}{7} \rho_0 \vsh^2,
\end{equation}
and its density to $7\rho_0$ (see eqs. \eqref{eq:rhops}, \eqref{eq:eps}). As the shocked fluid expands, it accelerates, converting its internal energy to kinetic energy. The final velocity, $\vt_f(r_0)$, of the fluid initially lying at $r_0$ is well approximated (for $\delta\ll1$) by $\vt_f(r_0)=f_\vt \vsh(r_0)$ with $f_\vt=2.16(2.04)$ for $n=3/2(3)$ \cite{MM99} where $\vsh$ is given by eq. \eqref{eq:vs_nonrel0}.

In what follows we label the shells by their Lagrangian coordinate, $\delta_m(r_0)$. The density and pressure evolution of a given shell, $p(\delta_m,t)$ and $\rho(\delta_m,t)$, is adiabatic,
\begin{equation}
    \label{eq:p_aftr_expns}
    p(\delta_m,t) = \bigg[ \frac{\rho(\delta_m,t)}{7\rho_0(\delta_m)} \bigg]^{4/3} p_0(\delta_m),
\end{equation}
as long as its optical depth is large (in which case the effects of photon diffusion may be neglected). Once a fluid shell expands to a radius significantly larger than $R_*$, its pressure drops well below $p_0$ and its velocity approaches the final velocity $\vt_f$. At this stage the shell's radius and density are given by
\begin{equation}
    \label{eq:r_prop_nonrel}
    r(\delta_m,t) = \vt_f(\delta_m) t,
\end{equation}
\begin{equation}
    \rho = -\frac{M}{4\pi r^2 t  }
    \left( \frac{d\vt_f}{d\delta_m} \right)^{-1} =
    \frac{n+1}{\beta_1\,n} \frac{ M}{4  \pi t^3 \vt_f^3 }\delta_m.
\end{equation}
where the last equality holds to lowest order in $\delta$. The resulting density profile is steep, $d\ln\rho/d\ln r=d\ln\rho/d\ln \vt_f=-3-(n+1)/\beta_1n\approx-10$.

\subsection{Luminosity and effective temperature: Constant opacity}
\label{ssec:const_opacity}

For a time and space independent opacity $\kappa$ (which applies, e.g., for opacity dominated by Thomson scattering with constant ionization), the optical depth of the plasma lying above the shell marked by $\delta_m$ is
\begin{eqnarray}
    \label{eq:optcl_dpth_nonrel}
    \tau(\delta_m,t) & \equiv & \int_{r(\delta_m,t)}^{\infty} {\rm d}r \kappa \rho(r,t)
        = \frac{\kappa M }{4\pi} \int_{0}^{\delta_m} \frac{{\rm d}\delta_m'}{r^2(\delta_m')}
       \nonumber \\ & = & \frac{1}{1+2 \beta_1 n/(1+n)} \frac{\kappa M \delta_m}{4\pi t^2 \vt_f^2(\delta_m)},
\end{eqnarray}
where the last equality holds to lowest order in $\delta$ when eq.~(\ref{eq:r_prop_nonrel}) is satisfied. We define the Lagrangian location of the photosphere, $\delta_{m,ph}$, by $\tau(\delta_m = \delta_{m,ph} ,t) = 1$. For $n = 3/2$ and $n = 3$ envelopes we find
\begin{eqnarray}
    \label{eq:ph_prop_nonrel}
    \delta_{m,ph}(t) &= 2.40 \times 10^{-3} f_{\rho}^{-0.12}
    \frac{ E_{51}^{0.81}}{(M/M_{\odot})^{1.63} \kappa^{0.81}_{0.34} }
    t_5^{1.63} \, (n=\frac{3}{2}),\nonumber\\
    \delta_{m,ph}(t) &= 2.62 \times 10^{-3} f_{\rho}^{-0.073}
    \frac{ E_{51}^{0.78}}{(M/M_{\odot})^{1.63} \kappa^{0.78}_{0.34} }
    t_5^{1.56} \, (n=3),
\end{eqnarray}
with corresponding photospheric radii
\begin{eqnarray}
\label{eq:non_rel_r_ph}
    r_{{\rm ph}}(t) &= 3.3 \times 10^{14} f_{\rho}^{-0.062}
    \frac{E_{51}^{0.41} \kappa_{0.34}^{0.093} }{(M/M_{\odot})^{0.31}}
    t_5^{0.81}{\rm cm} \, (n=\frac{3}{2}),\nonumber \\
    r_{{\rm ph}}(t) &= 3.3 \times 10^{14} f_{\rho}^{-0.036}
    \frac{E_{51}^{0.39} \kappa_{0.34}^{0.11} }{(M/M_{\odot})^{0.28}}
    t_5^{0.78}{\rm cm} \, (n=3),
\end{eqnarray}
with $t = 10^5\, t_5{\rm sec}$. Here, and in what follows, we use \cite{MM99} $ \beta_1 = 0.1909, f_v = 2.1649, {\rm and ~ }\Av = 0.7921 $ for $n=3/2$ and $\beta_1 = 0.1858, f_v = 2.0351, {\rm and ~}\Av = 0.8046 $ for $n=3$. Assuming that photon diffusion does not lead to significant deviations from the adiabatic evolution described by eq.~(\ref{eq:p_aftr_expns}), the effective temperature of the photosphere is given by
\begin{eqnarray}
    \label{eq:non_rel_T_ph}
    T_{{\rm ph}}(t) =1.6 \, f_{\rho}^{-0.037}
    \frac{E_{51}^{0.027} R_{*,13}^{1/4} }{(M/M_{\odot})^{0.054} \kappa^{0.28}_{0.34}}
    t_5^{-0.45} {\rm eV} \, (n=\frac{3}{2}),\nonumber \\
    T_{{\rm ph}}(t) =1.6 \, f_{\rho}^{-0.022}
    \frac{E_{51}^{0.016} R_{*,13}^{1/4} }{(M/M_{\odot})^{0.033} \kappa^{0.27}_{0.34}}
    t_5^{-0.47} {\rm eV} \, (n=3),
\end{eqnarray}
with $R_* = 10^{13} R_{*,13} {\rm cm} $. Note, that eq.~(\ref{eq:non_rel_T_ph}) corrects a typo (in the numerical coefficient) in eq.~(19) of \cite{WaxmanCampana07}. Approximating the luminosity by $L = 4\pi \sigma r_{{\rm ph}}^2 T_{{\rm ph}}^4$ we find
\begin{eqnarray}
    \label{eq:L_RBSG}
    L = 8.5\times 10^{42} \frac{E^{0.92}_{51} R_{*,13} }
    {f^{0.27}_{\rho} (M/M_{\odot})^{0.84}\kappa^{0.92}_{0.34} }t^{-0.16}_5 {\rm erg}\, {\rm s}^{-1}
     \, (n=\frac{3}{2}),\nonumber \\
    L = 9.9\times 10^{42} \frac{E^{0.85}_{51} R_{*,13} }
    {f^{0.16}_{\rho} (M/M_{\odot})^{0.69}\kappa^{0.85}_{0.34} }t^{-0.31}_5 {\rm erg}\, {\rm s}^{-1} \, (n=3).
\end{eqnarray}
The dependence on $n$ and on $f_{\rho}$ is weak.

Let us next examine the assumption, that photon diffusion does not lead to strong deviations from adiabatic expansion below the photosphere. In regions where the diffusion time is short compared to the expansion time, $t$, the luminosity carried by radiation, $L\propto r^2 dp/d\tau$, is expected to be independent of radius. The steep dependence of the density on radius, $d\ln\rho/d\ln r\sim-10$, then implies that the energy density in such regions roughly follows $p\propto\tau$, which is close to the adiabatic profiles derived in \S~\ref{ssec:hydro}, for which $p\propto \tau^{1.1}$ (for both $n=3,3/2$). Thus, we expect eqs.~(\ref{eq:non_rel_T_ph}) and~(\ref{eq:L_RBSG}) to provide reasonable approximations for the effective temperature and luminosity.
The validity of this conclusion may be tested by using the self-similar solutions of ref.~\cite{Chevalier92} for the diffusion of radiation in a constant opacity expanding envelope with power-law density and pressure profiles, $\rho\propto r^{-m}t^{m-3}$ and initial pressure $p\propto r^{-l}$, which yields a luminosity
\begin{eqnarray}
    \label{eq:L_Chev}
    L_c = 1.0\times 10^{43} \frac{E^{0.96}_{51} R_{*,13} }
    {f^{0.28}_{\rho} (M/M_{\odot})^{0.87}\kappa^{0.91}_{0.34} }t^{-0.17}_5 {\rm erg}\, {\rm s}^{-1}\, (n=\frac{3}{2}),\nonumber \\
    L_c = 9.6\times 10^{42} \frac{E^{0.91}_{51} R_{*,13} }
    {f^{0.17}_{\rho} (M/M_{\odot})^{0.74}\kappa^{0.82}_{0.34} }t^{-0.35}_5 {\rm erg}\, {\rm s}^{-1} \, (n=3).
\end{eqnarray}
($L$ derived in ref.~\cite{Chevalier92} (eq. 3.20) is different both in normalization and in scaling from those given in eq.~(\ref{eq:L_Chev}). This is due to some typographical errors in earlier eqs. of that paper. When corrected, in \cite{ChevalierFransson08}, the results obtained using the diffusion solutions are similar to those given here. See \cite{RW11} for details.) The parameter dependence of $L_c$ is similar to that obtained by the simple model described above, and the normalization of $L_c$, differs from that of eq.~(\ref{eq:L_RBSG}) by $\approx10\%$. It should be noted here that since the diffusion approximation breaks down near the photosphere, an exact solution requires using the transport equation.

\subsection{Opacity variation due to recombination}
\label{ssec:recombination}

The approximation of space and time independent opacity is justified at early times, when the envelope is highly ionized and the opacity is dominated by Thomson scattering. On a day time scale, the temperature of the expanding envelope drops to $\sim1$~eV, see eq.~(\ref{eq:non_rel_T_ph}). At this temperature, significant recombination may take place, especially for He dominated envelopes, leading to a significant modification of the opacity. The model presented in \S~\ref{ssec:const_opacity} is generalized in this section to include a more realistic description of the opacity. The deviation of the emitted spectrum from a black body spectrum, due to photon diffusion, is discussed in \S~\ref{ssec:Tc}.

Throughout this section, we use the density structure given by eq.~(\ref{eq:init_dnsty_prfl}) with $n=3$, as appropriate for radiative envelopes. As explained in the previous section, the results are not sensitive to the exact value of $n$.

In order to obtain a more accurate description of the early UV/O emission, we use the mean opacity provided in the OP project tables \cite{OPCD} (see ref.~\cite{RW11} for a brief discussion of the effect of line opacity enhancement due to velocity gradients). We replace eq.~(\ref{eq:optcl_dpth_nonrel}) with
\begin{equation}
\label{eq:tau_exact}
    \tau(\delta_{m},t)=
    \int_{r(\delta_{m},t)}^{\infty}{\rm d}r\rho\,\kappa[T(\delta_{m},t),\rho(\delta_{m},t)],
\end{equation}
where $\kappa(T,\rho)$ is the Rosseland mean of the opacity, and solve $\tau(\delta_m = \delta_{m,ph} ,t) = 1$ numerically for the location of the photosphere. In order to simplify the comparisons with the suggested analytical models, in the reminder of this section we shall take the ejecta properties in the limit of eq.~(\ref{eq:r_prop_nonrel}).

\subsubsection{Hydrogen envelopes}
\label{sssec:H_envelope}

For H dominated envelopes, the temperature of the photosphere calculated using the OP tables differs from that of eq.~(\ref{eq:non_rel_T_ph}) with $\kappa=0.34{\rm cm^2/g}$, corresponding to fully ionized 70:30 (by mass) H:He mixture, by less than 10\% for $T_{{\rm ph}}>1$~eV. At lower temperatures, the $\kappa=0.34{\rm cm^2/g}$ approximation leads to an underestimate of $T_{{\rm ph}}$, by $\approx20\%$ at 0.7~eV. This is due to the reduction in opacity accompanying H recombination. The reduced opacity implies that the photosphere penetrates deeper into the expanding envelope, to a region of higher temperature. The photospheric radius is not significantly affected and is well described by eq.~(\ref{eq:non_rel_r_ph}).

\subsubsection{He envelopes}
\label{sssec:He_envelope}

For He dominated envelopes, the constant opacity approximation does not provide an accurate description of $T_{{\rm ph}}$. We therefore replace eqs.~(\ref{eq:non_rel_r_ph}) and~(\ref{eq:non_rel_T_ph}) with an approximation, given in eqs.~(\ref{eq:EffModel-THe}) and~(\ref{eq:EffModel-r}), which takes into account the reduction of the opacity due to recombination, based on the numeric calculation. The approximation of eq.~(\ref{eq:EffModel-THe}) differs by less than 8\% from the result of a numerical calculation using the OP opacity tables down to $T_{{\rm ph}} \simeq 1$~eV. The temperature does not decrease significantly below $\simeq 1$~eV due to the rapid decrease in opacity below this temperature, which is caused by the nearly complete recombination.

On the time scale of interest, 1~hour~$\le t \le 1$~day, the photospheric temperature is in the energy range of 3eV~$\ge T \ge$~1eV. In this temperature range (and for the characteristic densities of the photosphere), the opacity may be crudely approximated by a broken power law,
\begin{equation}
\label{eq:approxHe_opacity}
\kappa = 0.085 \,({\rm cm^2/g})
\left\{
  \begin{array}{ll}
    (T/{\rm 1.07\, eV})^{0.88} , & \hbox{$T > {\rm 1.07\, eV}$;} \\
    (T/{\rm 1.07\, eV})^{10}, & \hbox{$T \leq {\rm 1.07\, eV}$.}
  \end{array}
\right.
\end{equation}
Using this opacity approximation, we find that eq.~(\ref{eq:non_rel_T_ph}) for the photospheric temperature is modified to
\begin{equation}
\label{eq:EffModel-THe}
T_{{\rm ph}}(t) =
\left\{
  \begin{array}{ll}
    1.33 {\rm eV} f_\rho^{-0.02}R_{*,12}^{0.20} t_5^{-0.38}, & \hbox{$T_{{\rm ph}} \geq {\rm 1.07\, eV}$;} \\
    {\rm 1.07\, eV} (t/t_b)^{-0.12},                         &  \hbox{$T_{{\rm ph}} < {\rm 1.07\, eV}$.}
  \end{array}
\right.
\end{equation}
Here, $R_*=10^{12}R_{*,12}$~cm and $t_b$ is the time at which $T_{{\rm ph}}={\rm 1.07\, eV}$, and we have neglected the dependence on $E$ and $M$, which is very weak. The photospheric radius, which is less sensitive to the opacity modification, is approximately given by
\begin{equation}
\label{eq:EffModel-r}
    r_{{\rm ph}}(t) = 2.8\times 10^{14} f_\rho^{-0.038} E_{51}^{0.39} (M/M_{\odot})^{-0.28} t_5^{0.75} {\rm cm}.
\end{equation}
Here we have neglected the dependence on $R_*$, which is weak. For $T_{{\rm ph}}>{\rm 1.07\, eV}$, the bolometric luminosity is given by
\begin{equation}
    \label{eq:L_He}
    L = 3.3\times 10^{42}
    \frac{E^{0.84}_{51} R_{*,12}^{0.85} }
    {f^{0.15}_{\rho} (M/M_{\odot})^{0.67} }t^{-0.03}_5 {\rm erg}\, {\rm s}^{-1}.
\end{equation}

The following comment is in place here. The strong reduction in opacity due to He recombination implies that the photosphere reaches deeper into the envelope, to larger values of $\delta_m$, where the initial density profile is no longer described by eq.~(\ref{eq:init_dnsty_prfl}) and the evolution of the ejecta is no longer given by the eqs. of \S~\ref{ssec:hydro}. This further complicates the model for the emission on these time scales.

\subsubsection{C/O and He-C/O envelopes}
\label{sssec:HeCO_envelope}

Finally, we consider in this section envelopes composed of a mixture of He and C/O. At the relevant temperature and density ranges, the C/O opacity is dominated by Thomson scattering of free electrons provided by these atoms, and is not very sensitive to the C:O ratio. Denoting by 1-Z the He mass fraction, the C/O contribution to the opacity may be crudely approximated, within the relevant temperature and density ranges, by
\begin{equation}
\label{eq:approxCO_opacity}
\kappa =  0.043\,\mbox{Z} (T/1\,{\rm eV})^{1.27}\,{\rm cm^2/g}.
\end{equation}
This approximation holds for a 1:1 C:O ratio. However, since the opacity is not strongly dependent on this ratio, $T_{{\rm ph}}$ obtained using eq.~(\ref{eq:approxCO_opacity}) holds for a wide range of C:O ratios (see discussion at the end of this subsection). At the regime where the opacity is dominated by C/O, eq.~(\ref{eq:non_rel_T_ph}) is modified to
\begin{equation}
\label{eq:EffModel-TCO}
T_{{\rm ph}}(t) = 1.5 {\rm eV} f_\rho^{-0.017} \mbox{Z}^{-0.2}  R_{*,12}^{0.19} t_5^{-0.35}.
\end{equation}

In the absence of He, i.e. for $\mbox{Z}=1$, $T_{{\rm ph}}$ is simply given by eq.~(\ref{eq:EffModel-TCO}). For a mixture of He-C/O, $\mbox{Z}<1$, $T_{{\rm ph}}$ may be obtained as follows. At high temperature, where He is still ionized, the He and C/O opacities are not very different and $T_{{\rm ph}}$ obtained for a He envelope, eq.~(\ref{eq:EffModel-THe}), is similar to that obtained for a C/O envelope, eq.~(\ref{eq:EffModel-TCO}). At such temperatures, we may use eq.~(\ref{eq:EffModel-THe}) for an envelope containing mostly He, and  eq.~(\ref{eq:EffModel-TCO}) with $\mbox{Z}=1$ for an envelope containing mostly C/O (a more accurate description of the Z-dependence may be straightforwardly obtained by an interpolation between the two equations). At lower temperature, the He recombines and the opacity is dominated by C/O. At these temperatures, $T_{{\rm ph}}$ is given by eq.~(\ref{eq:EffModel-TCO}) with the appropriate value of $\mbox{Z}$. The transition temperature is given by
\begin{equation}
\label{eq:EffModel-TransT}
    T_{\rm He-C/O} = 1\, \mbox{Z}^{0.1} {\rm eV}.
\end{equation}
The photospheric radius, which is less sensitive to the opacity variations, is well approximated by eq.~(\ref{eq:EffModel-r}).
At the stage where the opacity is dominated by C/O, the bolometric luminosity is given by
\begin{equation}
    \label{eq:L_CO}
    L = 4.7\times 10^{42} \frac{E^{0.83}_{51} R_{*,12}^{0.8} }
    {f^{0.14}_{\rho} \mbox{Z}^{0.63}(M/M_{\odot})^{0.67} }t^{0.07}_5 {\rm erg}\, {\rm s}^{-1}.
\end{equation}

For C/O envelopes, the analytic approximation for $T_{{\rm ph}}$ derived above,  eq.~(\ref{eq:EffModel-TCO}), differs by less than 6\% from the result of a numerical calculation using the OP opacity tables down to $T_{\rm ph} \simeq0.5$~eV. For Z in the range $0.7> \mbox{Z}>0.3$, the approximations obtained by using eqs.~(\ref{eq:EffModel-THe}) and~(\ref{eq:EffModel-TCO}) with a transition temperature given by eq.~(\ref{eq:EffModel-TransT}) hold to better than $\approx10\%$ down to $T_{{\rm ph}} \simeq 0.8$~eV.

\subsection{Color temperature}
\label{ssec:Tc}

We have shown in \S~\ref{ssec:const_opacity} that photon diffusion is not expected to significantly affect the luminosity. Such diffusion may, however, modify the spectrum of the emitted radiation. We discuss below in some detail the expected modification of the spectrum.

For the purpose of this discussion, it is useful to define the "thermalization depth", $r_{{\rm ther}}$, and the "diffusion depth", $r_{{\rm diff}}$. $r_{{\rm ther}}(t)<r_{{\rm ph}}(t)$ is defined as the radius at which photons that reach $r_{{\rm ph}}(t)$ at $t$ "thermalize", i.e. the radius from which photons may reach the photosphere without being absorbed on the way. This radius may be estimated as the radius for which $\tau_{{\rm sct}} \tau_{{\rm abs}} \approx 1 $ \cite{Mihalas84}, where $\tau_{{\rm sct}}$ and $\tau_{{\rm abs}}$ are the optical depths for scattering and absorption provided by plasma lying at $r>r_{{\rm ther}}(t)$. $r_{{\rm ther}}$ is thus approximately given by
\begin{equation}
\label{eq:r_thrm}
    3 (r_{{\rm ther}}-r_{{\rm ph}})^2 \kappa_{{\rm sct}}(r_{{\rm {\rm ther}}}) \kappa_{{\rm abs}}(r_{{\rm {\rm ther}}}) \rho^2(r_{{\rm {\rm ther}}}) = 1,
\end{equation}
where $\kappa_{{\rm {\rm sct}}} $ and $\kappa_{{\rm abs}}$ are the scattering and absorption opacities respectively (typically, the opacity is dominated by electron scattering). $r_{{\rm diff}}$ is defined as the radius (below the photosphere) from which photons may escape (i.e. reach the photosphere) over a dynamical time (i.e. over $t$, the time scale for significant expansion). We approximate $r_{{\rm diff}}$ by
\begin{equation}
\label{eq:r_diff}
    r_{{\rm ph}} = r_{{\rm diff}}+\sqrt{c \, t/3 \kappa_{{\rm {\rm sct}}}(r_{{\rm diff}}) \rho(r_{{\rm diff}})},
\end{equation}
where $c$ is the speed of light.

For $r_{{\rm diff}} < r_{{\rm {\rm ther}}}$, photons of characteristic energy $3T(r_{{\rm {\rm ther}}},t)>3T_{{\rm ph}}$ will reach the photosphere, while for $r_{{\rm {\rm ther}}} < r_{{\rm diff}}$ photons of characteristic energy $3T(r_{{\rm diff}},t)>3T_{{\rm ph}}$ will reach the photosphere. Thus, the spectrum will be modified from a black body at $T_{{\rm ph}}$ and its color temperature, $T_{\rm col}$ (with specific intensity peaking at $3T_{{\rm col}}$) will be $T_{{\rm col}}>T_{{\rm ph}}$.

Approximating $T_{{\rm col}}=T(r_{{\rm {\rm ther}}})$ for $r_{{\rm diff}} < r_{{\rm {\rm ther}}}$ and $T_{{\rm col}}=T(r_{{\rm diff}})$ for $r_{{\rm diff}} > r_{{\rm {\rm ther}}}$, the ratio $T_{{\rm col}}/T_{{\rm ph}}$ was calculated in ref.~\cite{RW11} assuming that the scattering opacity is dominated by Thomson scattering of free electrons (with density provided by the OP tables), and estimating $\kappa_{{\rm abs}}=\kappa-\kappa_{{\rm {\rm sct}}}$ (recall that $\kappa$ is the Rosseland mean of the opacity). It would have been more accurate to use an average of the absorptive opacities over the relevant wavebands, which are not provided by the OP table. However, since the dependence of the color temperature on the absorptive opacity is weak, $T_{{\rm col}} \propto \kappa_{{\rm {\rm abs}}}^{(-1/8)}$, the corrections are not expected to be large.

Under the above assumptions, $T_{{\rm col}}/T_{{\rm ph}}$ is approximately given, for $t\le1$~day, by
\begin{equation}
    \label{eq:f_T}
    f_T \equiv T_{{\rm {\rm col}}}/T_{{\rm {\rm ph}}} \approx 1.2\quad.
\end{equation}

Using eq.~(\ref{eq:f_T}) with eqs.~(\ref{eq:non_rel_T_ph}),~(\ref{eq:EffModel-THe}) and~(\ref{eq:EffModel-TCO}) for the photospheric (effective) temperature, the progenitor radius may be approximately inferred from the color temperature by \begin{equation}
    \label{eq:R_H}
    R_{*} \approx  0.70\times10^{12} \left[ \frac{T_{{\rm col}}}{(f_T/1.2) {\rm eV}} \right]^{4}t_5^{1.9}
    f_{\rho}^{0.1}\,\rm cm
\end{equation}
for H envelopes,
\begin{equation}
    \label{eq:R_He}
    R_{*} \approx  1.2\times 10^{11} \left[ \frac{T_{{\rm col}}}{(f_T/1.2) {\rm eV}} \right]^{4.9}   t_5^{1.9} f_{\rho}^{0.1}\, {\rm cm}
\end{equation}
for He envelopes with $T>{\rm 1.07\, eV}$, and
\begin{equation}
    \label{eq:R_He-CO}
    R_{*} \approx  0.58 \times 10^{11} \left[ \frac{T_{{\rm col}}}{(f_T/1.2) {\rm eV}} \right]^{5.3}   t_5^{1.9} f_{\rho}^{0.1} \mbox{Z} \,{\rm cm}
\end{equation}
for He-C/O envelopes when the C/O opacity dominates (the transition temperature is given in eq.~(\ref{eq:EffModel-TransT})).

\subsection{Removing the effect of reddening}
\label{ssec:reddening}

We show in this section that the effects of reddening on the observed UV/O signal may be removed using the UV/O light curves. This is particularly important for inferring $R_*$, since $R_*\propto T_{\rm col}^\alpha$ with $4\le\alpha\le5$ (see eqs.~(\ref{eq:R_H})--(\ref{eq:R_He-CO})).

The model specific intensity, $f_\lambda$, is given by
\begin{equation}\label{eq:f_lambda}
    f_\lambda(\lambda,t)=\left(\frac{r_{\rm ph}}{D}\right)^{2}\sigma T_{\rm ph}^4 \frac{T_{\rm col}}{hc}
    g_{BB}(hc/\lambda T_{\rm col}) e^{-\tau_\lambda},
\end{equation}
where
\begin{equation}\label{eq:g_BB}
    g_{BB}(x)=\frac{15}{\pi^4} \frac{x^5}{e^x-1},
\end{equation}
$D$ is the distance to the source, and $\tau_\lambda$ is the extinction optical depth at $\lambda$.
Let us define $t_\lambda(t,\lambda)$ by
\begin{equation}\label{eq:t_lambda}
    \lambda T_{\rm col}[t=t_\lambda(t,\lambda)]=\lambda_0 T_{\rm col}(t),
\end{equation}
for some chosen $\lambda_0$. With this definition, the scaled light curves,
\begin{eqnarray}
\label{eq:rescaled_flux}
    \tilde{f}_{\lambda}[\lambda,t_\lambda(t,\lambda)]\equiv
    \left[\frac{D}{r_{\rm ph}(t_\lambda)}\right]^{2}
    \left[\frac{T_{\rm col}(t_\lambda)}{T_{\rm ph}(t_\lambda)}\right]^{4}
    \left[\frac{T_{0}}{T_{\rm col}(t_\lambda)}\right]^{5}
    \times f_{\lambda}\left(\lambda,t_\lambda\right)
\end{eqnarray}
(where $T_0$ is an arbitrary constant) are predicted to be the same for any $\lambda$ up to a factor $e^{-\tau_\lambda}$,
\begin{equation}\label{eq:scaled_model}
    \tilde{f}_{\lambda}[\lambda,t_\lambda(t,\lambda)]=\sigma T_0^4 \frac{T_0}{hc}g_{BB}[hc/\lambda_0 T_{\rm col}(t)]\times e^{-\tau_\lambda}.
\end{equation}

Let us consider now how the scalings defined above allow one to determine the relative extinction in cases where the model parameters $\{E,M,R_*\}$ are unknown, and hence $\{T_{\rm col},T_{\rm ph},r_{\rm ph}\}(t)$, which define the scalings, are also unknown. For simplicity, let us first consider the case where the time dependence of the photospheric radius and temperature are well approximated by power-laws,
\begin{equation}
\label{eq:power-laws}
    r_{\rm ph}\propto t^{\alpha_r}, \quad T_{\rm ph}\propto t^{-\alpha_T},
\end{equation}
and the ratio $T_{\rm col}/T_{\rm ph}$ is independent of time. This is a good approximation for the time dependence of $r_{\rm ph}$ in general, and for the time dependence of $T_{\rm col}$ and $T_{\rm ph}$ for $T_{\rm ph}>1$~eV (see eqs.~\ref{eq:non_rel_r_ph},~\ref{eq:non_rel_T_ph},~\ref{eq:EffModel-THe},~\ref{eq:EffModel-r},~\ref{eq:EffModel-TCO}, and~\ref{eq:f_T}). In this case eq.~(\ref{eq:t_lambda}) gives
\begin{equation}\label{eq:t_scale}
    t_\lambda(t,\lambda)=\left(\frac{\lambda}{\lambda_0}\right)^{1/\alpha_T}t\,,
\end{equation}
and eq.~(\ref{eq:rescaled_flux}) may be written as
\begin{eqnarray}\label{eq:f_scale}
    \tilde{f}_{\lambda}[\lambda,t_\lambda(t,\lambda)]&=&{\rm Const.}\times
    \left(\frac{\lambda}{\lambda_0}\right)^{(-2\alpha_r+5\alpha_T)/\alpha_T} t^{-2\alpha_r+5\alpha_T}\nonumber\\
    &\times& f_{\lambda}\left[\lambda,\left(\frac{\lambda}{\lambda_0}\right)^{1/\alpha_T}t\right]\,.
\end{eqnarray}
The value of the constant that appears in eq.~(\ref{eq:f_scale}), for which the normalization of $\tilde{f}_{\lambda}$ is that given by eq.~(\ref{eq:scaled_model}), is not known, since it depends on the model parameters $\{E,M,R_*\}$. However, for any choice of the value of the constant, $\tilde{f}_{\lambda}$ defined by eq.~(\ref{eq:f_scale}) is predicted by the model to be given by eq.~(\ref{eq:scaled_model}) up to a wavelength independent multiplicative factor. Thus, the ratio of the scaled fluxes defined in eq.~(\ref{eq:f_scale}) determines the relative extinction,
\begin{equation}\label{eq:reddening}
    \frac{\tilde{f}_{\lambda}[\lambda_1,t_\lambda(t,\lambda_1)]}
    {\tilde{f}_{\lambda}[\lambda_2,t_\lambda(t,\lambda_2)]}= e^{\tau_{\lambda_2}-\tau_{\lambda_1}}\, .
\end{equation}

Let us consider next the case where the time dependence of $T_{\rm col}$ and $T_{\rm ph}$ is not a simple power-law. $T_{\rm col}$ and $T_{\rm ph}$ are determined by the composition and progenitor radius $R_*$, and are nearly independent of $E$ and $M$. Adopting some value of $R_*$, eq.~(\ref{eq:t_lambda}) may be solved for $t_\lambda(t,\lambda;R_*)$ and eq.~(\ref{eq:rescaled_flux}) may be written as
\begin{eqnarray}\label{eq:f_scale_p}
    \tilde{f}_{\lambda}[\lambda,t_\lambda(t,\lambda;R_*)]={\rm Const.}\times t^{-2\alpha_r}
    T_{\rm ph}(t_\lambda)^{-4}T_{\rm col}(t_\lambda)^{-1}
    \times f_{\lambda}\left(\lambda,t_\lambda\right).
\end{eqnarray}
The model predicts therefore that scaling the observed flux densities using the correct value of $R_*$, the observed light curves at all wavelengths should be given by eq.~(\ref{eq:scaled_model}), up to a multiplicative wavelength independent constant. For this value of $R_*$, the ratio of the scaled fluxes at different wavelengths is independent of $t$ and given by eq.~(\ref{eq:reddening}). The value of $R_*$ may be therefore determined by requiring the ratios of scaled fluxes to be time independent, and the relative extinction may then be inferred from eq.~(\ref{eq:reddening}).

\section{Using stellar-surface breakout and cooling emission observations to constrain progenitor and explosion parameters}
\label{sec:ObsTheory}

\subsection{Breakout burst}
\label{ssec:ObsTheBO}
The properties of the breakout burst depend on the radius of the star $R_*$, the velocity of the breakout $\vBO$ and to a lesser extent on the density at breakout $\rhoBO$, and thus can be used to constrain these parameters. These in turn can be used to constrain global properties of the ejecta (in particular $\vt_*=\sqrt{E/M}$) through the approximate hydrodynamic relations \eqref{eq:MVParam_v} and \eqref{eq:PlanarMVParam_rho}.   The properties are insensitive to the density profile.

The first detected bursts will likely have a limited amount of photons. Even with a few tens of photons, the total energy in the burst $\EBO$ and the frequency $\nu_{\rm peak}$ at which the fluence peaks (in terms of energy per logarithmic frequency) are likely to be reliably measurable and are expressed in Eq. \eqref{eq:PlanarEinfty} and \eqref{eq:PlanarTVnumeric}. These quantities are insensitive to slight deviations from spherical symmetry. The total energy in the burst, is approximately equal to 2 times the peak fluence per logarithmic frequency, (Eq. \eqref{eq:PlanarnuEnupeak}) and thus measuring the peak is sufficient for both quantities. The low dependence on the breakout density, implies that the measurement these quantities is sufficient to obtain an approximate measurement of the progenitor radius (velocity from \eqref{eq:PlanarTVnumeric} and then radius from \eqref{eq:PlanarEinfty}). The timescale of the burst is also likely measurable and provides an upper limit for the radius since the burst duration must be greater than the light crossing time $R/c$.

If the light curve can be reliably determined and exact spherical symmetry is assumed, the light curve can provide an independent measurement of $R$. In this case, the luminosity and temperature evolution can be calculated accurately \cite{Sapir11,Katz12,SKWspec13,SapirHalbertal14}. In particular, the peak bolometric luminosity is approximately given by equation 39 (using equations 5 and 32) in \cite{Katz12}. The amplitude of the bolometric luminosities at intermediate times $R/c\lesssim t \lesssim R/4\vBO$ is insensitive to deviations from spherical symmetry and is given by equation \eqref{eq:PlanarELApprox}. If measured (overcoming the likely challenge of ISM absorption given the decreasing temperatures) the luminosity at such times may provide an independent robust measurement of the radius.

For large RSG progenitors, where the emitted energy is largest, the expected temperatures, of tens of eV, are such that most of the radiation is expected to be absorbed in the ISM. Very small progenitors such as Wolf-Rayet (WR) stars are expected to produce multi-keV  photons but with a very small output $E\lesssim 2\times 10^{44}R_{11}^{2}\rm erg $ (see Eq. \ref{eq:PlanarEinfty}, assuming non-relativistic breakout $\vBO<10^{10}\rm cm/s$). The most promising progenitors for observable breakout burst are intermediate size BSG progenitors were $\sim keV$ photons may be produced with observable outputs (see \S\ref{ssec:surface_obs}) emission with smaller outputs.

\subsection{Post breakout cooling}
\label{ssec:ObsTheCool}

The post breakout cooling emission is nearly independent of the density structure of the outer envelope as long as the emission is dominated by $\delta\ll1$ shells, see eq.~(\ref{eq:t_valid}). In this limit, the luminosity $L$ and the effective temperature $T$ of the emitted radiation are determined by $R_*$, $E/M$ and $\kappa$. $T_{\rm ph}(t)$ and $L(t)$ are given by eqs.~(\ref{eq:non_rel_T_ph}) and~(\ref{eq:L_RBSG}) for H dominated envelopes, by eqs.~(\ref{eq:EffModel-THe}) and~(\ref{eq:L_He}) for He envelopes, and by eqs.~(\ref{eq:EffModel-TCO}-\ref{eq:L_CO}) for He:C/O envelopes. The ratio of color to photospheric temperature is approximately $T_{\rm col}/T_{\rm ph}=1.2$ (see \S~\ref{ssec:Tc}), and the spectral luminosity per unit wavelength $\lambda$ may be approximated by
\begin{equation}\label{eq:L_spec}
  L_\lambda(t)= L(t)\frac{T_{\rm col}}{hc}g_{\rm BB}(hc/\lambda T_{\rm col}),
\end{equation}
where $g_{\rm BB}$ is the normalized Planck function,
\begin{equation}\label{eq:gBB}
  g_{\rm BB}(x)= \frac{15}{\pi^4}\frac{x^5}{e^x-1}.
\end{equation}
The very weak dependence of $T$ on $E/M$ implies that $R_*/\kappa$ may be inferred from a measurement of the color temperature, see eqs.~(\ref{eq:R_H}--\ref{eq:R_He-CO}), and that $E/M$ may be inferred from the bolometric luminosity $L$. Several points should be carefully taken into consideration when inferring $R_*$ and $E/M$ from observations.

Let us first consider the effects of reddening. The strong dependence of $R_*$ on $T$, see eqs.~(\ref{eq:R_H})--(\ref{eq:R_He-CO}), implies that an estimate of $R_*$ based on a determination of $T$ from the spectrum observed at a given time $t$ would be sensitive to reddening. On the other hand, the value of $T$ at a given $t$ may be inferred from the light curve at a given wavelength, $L_\lambda(t)$, since the time $t_\lambda$ at which $L_\lambda(t)$ reaches its maximum is approximately the time at which $T$ crosses $hc/4\lambda$ (the exact value is determined from the model's $L(t)$ and $T(t)$). Moreover, inferring $T$ from the shape of the light curve allows one to infer the reddening by comparing the fluxes observed at different wavelengths, as explained in some detail in \S~\ref{ssec:reddening}. Finally, in order to infer $E/M$ from the luminosity $L$, the absolute value of the extinction should be determined. This cannot be done without some assumption on the relation between the (measured) reddening and the absolute extinction.

Next, temporal and spatial opacity variations due to recombination should be taken into account for envelopes that are not H dominated, see eqs.~(\ref{eq:approxHe_opacity}) and~(\ref{eq:approxCO_opacity}). The opacity variations lead to a modification of the model light curves, as described in \S~\ref{sssec:He_envelope} and \S~\ref{sssec:HeCO_envelope}. A detailed measurement of the light curve may thus constrain the composition of the outer envelope.

Finally, it should be noted that the model described here is valid only at early times, of order a few days, during which the emission is dominated by $\delta\ll1$ shells (see eq.~(\ref{eq:t_valid})) and $T\gtrsim1$~eV (see eqs.~(\ref{eq:non_rel_T_ph}),~(\ref{eq:EffModel-THe}),~(\ref{eq:EffModel-TCO})). At later time, as the photosphere penetrates to larger $\delta$ values, the evolution is no longer described by the simple self-similar solution given here and depends on the detailed structure of the ejecta. Moreover, as $T$ drops below 0.7~eV for H dominated envelopes, or 1~eV for He dominated envelopes, recombination leads to a strong decrease of the opacity with decreasing temperature. At this stage the photosphere penetrates deep into the ejecta, to a depth where the temperature is sufficiently high to maintain significant ionization and large opacity. This enhances the dependence on the details of the envelope structure and implies that detailed radiation transfer models are required to describe the emission (our simple approximations for the opacity no longer hold). An accurate and robust determination of $R_*$, and hence of $E/M$, requires an accurate determination of $T$, at times when $T$ depends mainly on $R_*$, see eqs.~(\ref{eq:R_H})--(\ref{eq:R_He-CO}), and independent of the details of the ejecta structure, i.e. at $T\gtrsim1$~eV. An accurate determination of $T$ requires one to observe at $\lambda<hc/4T=0.3(T/1{\rm eV})^{-1}\mu$, in order to identify the peak in the light curve (or the spectral peak if extinction effects can be reliably removed). UV observations are thus required for a robust and accurate determination of $R_*$ and $E/M$.

An elaborate example of the application of the method described above for inferring progenitor parameters may be found in ref.~\cite{RW11}. Analysis of the early UV/O observations of the type Ib SN2008D led to a determination of the progenitor's radius, $R_*\approx10^{11}$~cm, which cannot be directly inferred from later time observations, of $E/M$, $E_{51}/(M/M_\odot)\approx0.8$, of the reddening, $E(B-V)=0.6$, and to an indication that the He envelope of SN2008D contained a significant C/O fraction. The inferred values of $E/M$ and of the reddening, as well as the inferred presence of C/O, is consistent with later observations of the main SN light curve \cite{Mazzali08Sci,Tanaka09,Modjaz09}. The inferred radius constrains progenitor models, and is consistent with the calculations of ref.~\cite{Tanaka09}.

The example of SN2008D demonstrates the importance of including the time dependence of the opacity, using eqs.~(\ref{eq:EffModel-THe})--(\ref{eq:L_He}) instead of eqs.~(\ref{eq:non_rel_T_ph}) and~(\ref{eq:L_RBSG}), which are valid for constant opacity (H dominated) envelopes. It also demonstrates the importance of deviations from the solution presented here at late times, see eq.~(\ref{eq:t_valid}), when $\delta_{\rm ph}\ll1$ no longer holds. In the analysis of SN2008D given in ref.~\cite{RW11}, the model presented above was extended to large values of $\delta$ in order to extend the model predictions to $t\sim2$~d.

Analyses that do not include the opacity variation with time and extend the analytic model beyond the limit of eq.~(\ref{eq:t_valid}) (e.g.~\cite{ChevalierFransson08,Bersten_08D_13}) would find model predictions that are inconsistent with observations and with the results of ref.~\cite{RW11} (e.g. compare fig.~7 of ref.~\cite{Bersten_08D_13} with fig.~10 of ref.~\cite{RW11}).

\section{Open theoretical issues}
\label{sec:discussion}

We discuss in this section two topics, which are currently under vigorous theoretical investigation, and for which complete (near) exact solutions, as described in the preceding sections for stellar surface breakouts, are not yet available: breakouts from extended circumstellar media (\S~\ref{ssec:CSM}), and relativistic breakouts (\S~\ref{ssec:relativistic}). We outline the main open questions and the gaps in the theoretical analyses that need to be closed.

\subsection{Breakouts from an extended circumstellar medium}
\label{ssec:CSM}

In cases where the SN progenitor is surrounded by extended circumstellar medium (CSM), with optical depth exceeding $c/\vt_{\rm bo,*}$ where $\vt_{\rm bo,*}$ is the breakout velocity from the stellar surface, the RMS continues to propagate into the CSM, and breakout occurs at a radius $R_{\rm bo}>R_*$, at which the optical depth of the overlying material drops below $c/\vsh$. The presence of such extended CSM may be the result of a massive wind or of the ejection of outer envelope shells prior to the SN explosion. The dynamics of the RMS depositing energy into the CSM depends on the CSM density structure. One may consider, for example, a blast-wave driven by an ejecta expanding through a continuous wind, or a blast wave generated by the collision of an expanding ejecta with a detached CSM shell.

The characteristic duration of the pulse is $t_{\rm bo}\sim\tau R_{\rm bo}/c\sim R_{\rm bo}/\vt_{\rm bo}$, where $\vt_{\rm bo}$ is the breakout velocity in the CSM, and the characteristic luminosity may be estimated as $M \vt_{\rm bo}^2/t_{\rm bo}$ with $M$, the shocked CSM mass, estimated from from $\tau\sim \kappa M/4\pi R_{\rm bo}^2\sim c/\vt_{\rm bo}$ (e.g. \cite{OfekWind10,ChevalierIrwin11}),
\begin{equation}\label{eq:CSM_scales}
    t_{\rm bo}\sim 10^{5}\frac{R_{\rm bo,14}}{\vt_{\rm bo,9}}{\rm s},\quad
    L_{\rm bo}\sim\frac{4\pi c}{\kappa}R_{\rm bo}\vt_{\rm bo}^2\sim 10^{44}R_{\rm bo,14}\vt_{\rm bo,9}^2{\rm erg/s},
\end{equation}
where $R_{\rm bo}=10^{14}R_{\rm bo,14}$~cm, $\vt_{\rm bo}=10^{9}\vt_{\rm bo,9}{\rm cm/s}$. In the case of an extended wind the time scale may be increased by a factor $\sim\log(c/\vsh)$, and the luminosity scale may be correspondingly reduced, due to photon diffusion at $\tau<c/\vsh$. The spectrum of the emitted radiation is difficult to calculate, due to reasons detailed below.
In general, as the shock wave is converted from an RMS to a collisionless shock at breakout \cite{KSW11,KSW12}, a transition from UV dominated spectra to X-ray dominated spectra is expected. For fast breakouts, $\vt/c>0.1$, significant X-ray emission is expected at breakout, while for slower breakouts X-ray emission is expected to become dominant at later times (e.g. \cite{KSW11,Balberg11,ChevalierIrwin11,ChevalierIrwin12,SvirskiWind12,SvirskiNakar14}).
High energy photon ($h\nu\gg m_ec^2$) and neutrino (multi-TeV) emission is also expected, and may be detectable by existing telescopes, due to the acceleration of electrons and protons to high energy at the collisionless shock \cite{KSW11,Murase11CSM_TeV,KSW12,Murase14CSM_TeV,Kowalski15NuAstro,LAT-PTF15CSM,Zirakashvili15IInNu,Zirakashvili15aIInNU}.

CSM breakouts may thus produce large luminosities, due to conversion of kinetic energy to thermal energy at large radius, in which case adiabatic expansion losses are greatly reduced. This requires a significant optical depth at large radii. For a steady wind with constant mass loss, $\dot{M}$, and velocity, $\vt_w$, the requirement is $\dot{M}\sim(4\pi c/\kappa)(\vt_w/\vt_{\rm bo})R_{\rm bo}\sim 10^{-3}(\vt_w/10^{-3}\vt_{\rm bo})R_{\rm bo,14}M_\odot/{\rm yr}$.

There is significant and increasing evidence that large mass loss episodes closely preceding the stellar explosion are not uncommon (see \S~\ref{ssec:CSM_obs} for more details). The early lightcurves of SNe of type IIn are consistent with being generated by wind breakouts \cite{Ofek14IIn-wind,Drout14Wind,Gezari15Wind} with inferred mass loss rates $>10^{-3}M_\odot/{\rm yr}$ (e.g. \cite{Kiewe12Mdot,Taddia13Mdot,Fransson14Mdot,Ofek14extremeMdot,Moriya14Mdot,Moriya14Mdot2}), far exceeding the rates expected for line-driven winds (see \cite{Langer12Rev} for review). The detection of pre-SN "precursors" in several (mostly IIn) SNe \cite{Foley07precursor,Pastorello07precursor,Ofek13precursor,Pastorello13precursor,Fraser13precursor,Mauerhan13precursor,Corsi14precursor,Mauerhan14precursor}, and the indication that such precursors are common for IIn SNe on a month time scale preceding the explosion \cite{Ofek14Precursors}, provide independent evidence for intense mass loss episodes in many SN progenitors shortly before the explosion. Finally, the strong emission of super-luminous SNe (of type II/I(c)) may be interpreted as due to breakouts from extended CSM with (shocked) mass comparable to the SN ejecta mass \cite{Quimby07SL05ap,Smith07SL06gy,SmithMcCray07SL-CSM,Woosley07PIMdot,Miller09SL08es,OfekWind10,Smith10SL-CSM,Moriya11,Balberg11,ChevalierIrwin11,Chatzopoulos12,GinzburgBalberg12,Moriya13SL}.

We do not elaborate here on the theory of CSM breakouts. This is due to the fact that a complete theoretical analysis, as the one presented in the preceding sections for the the stellar surface breakout/post breakout cooling problem, is not yet available for CSM breakouts. The hydrodynamics of CSM breakouts and the resulting bolometric light curves have been thoroughly studied using analytic, semi-analytic and numeric methods for various CSM structures (e.g. \cite{OfekWind10,ChevalierIrwin11,Balberg11,Moriya11,Moriya13,SvirskiWind12,SvirskiNakar14,Chatzopoulos12,Chatzopoulos13,GinzburgBalberg12,GinzburgBalberg14,NakarPiro14,Piro15}). The study of the resulting spectra is on going and not yet complete. Analytic analyses provide heuristic qualitative descriptions of the expected spectra or characteristic radiation temperature (e.g. \cite{OfekWind10,Balberg11,KSW11,ChevalierIrwin12,SvirskiWind12,SvirskiNakar14}), while numeric analyses (e.g. \cite{Moriya11,KasenNumeric11,GinzburgBalberg12,Moriya13,KasenNumeric15}) do not yet include all the relevant physical process, as discussed in some detail below. The theoretical uncertainties affect the analysis of observations, as discussed in
\S~\ref{ssec:CSM_obs}.

The main challenge, that theoretical analyses of the CSM breakout spectra face, is related to the fact that the plasma in the shock transition region, where most of the radiation is generated, is not in general in thermal equilibrium, combined with the fact that the shock wave changes its structure in a complicated manner on a time scale comparable to the dynamical time scale, $R/\vsh$. In the stellar surface breakout situation, the mass lying ahead of the shock at breakout is small, and the escaping radiation is capable of accelerating the overlying shells to high velocity, followed by a phase of free expansion (see \S~\ref{sec:Planar}). In the CSM breakout case, the mass lying ahead of the shock at breakout may be large, and the radiation escaping from the RMS may not carry sufficient momentum to accelerate the overlying shells to high velocity. As a results, the faster inner shells drive a collisionless shock through the overlying CSM \cite{KSW12}. The electron spectrum at the shock transition, which determines the emitted radiation spectrum, is determined by interaction of the accelerated electrons with the ambient radiation, which is initially dominated by the RMS generated radiation diffusing outward, and later by Bremmsstrahlung emission at the shock transition \cite{SvirskiWind12,ChevalierIrwin11,ChevalierIrwin12}. An accurate, self consistent description of the evolution of the shock structure and of the emitted radiation is not yet available though progress has been made \cite{Balberg11,ChevalierIrwin11,ChevalierIrwin12,SvirskiWind12,SvirskiNakar14}. We briefly discuss below the implied challenges to analytic and numeric calculations.

Analytic analyses are challenged by several major factors.
\begin{itemize}
  \item The non-steady nature of the shock structure at breakout implies that solutions of steady shock structure are not applicable (For example, for stellar surface breakout, the breakout radiation temperature is 2--5 times lower than given by a steady RMS with $\vt=\vt_{\rm bo}$ \cite{SKWspec13}). The situation is more complicated than in the stellar surface case, due to the RMS-collisionless transition.
  \item The diversity and complexity of the spatial density profiles limit the usefulness of analytic (e.g. self-similar) solutions for the hydrodynamics.
  \item The formation of a collisionless shock leads to the generation of non-thermal high energy particles and photons. The complicated interaction of radiation with high energy electrons and the resulting complicated spectra are difficult to describe analytically.
\end{itemize}
Numeric analyses are challenged by several major factors as well.
\begin{itemize}
 \item The formation of a collisionless shock and the generation of non-thermal high energy particles and photons is difficult to include, and is not included in current numerical calculations.
 \item Inelastic Compton scattering plays a crucial role in determining the electron temperature and photon spectrum. This is quite challenging to include in radiation-hydro codes, and is not included in current numerical calculations (e.g. \cite{Moriya11,KasenNumeric11,GinzburgBalberg12}; see discussion in \cite{KasenNumeric11}). A Monte-Carlo algorithm, that in principle could accommodate a description of inelastic Compton scattering, was recently described in \cite{KasenNumeric15}, but a calculation including inelastic Compton scattering was not implemented.
  \item The cooling time of shock-heated electrons is much smaller than the dynamical time, which implies that in order to correctly determine their temperature (and the emitted radiation spectrum) a challengingly high resolution is required.
  \item The rapid electron cooling also leads, at high velocity, to a separation between the electron and the proton temperatures, which is not included in current numerical calculations.
\end{itemize}

\subsection{Relativistic breakouts}
\label{ssec:relativistic}
Several complications arise if the shock reaches mildly or ultra-relativistic velocities as it approaches the surface \cite{Weaver76,Levinson08,Katz10,Budnik10,NakarSari12} . The modifications required to the analysis presented in section \ref{sec:Planar} can be separated to those that are related to the high expected temperatures, which exceed tens of keV for $\vsh\gtrsim0.2 c$ (see eq. \eqref{eq:PlanarVTanalytic}), and to modifications due to the high velocities of the plasma. The high temperatures give rise to the following corrections and new effects:
\begin{itemize}
\item A significant amount of electron-positron pairs may be produced, increasing the optical depth of a given fluid element by orders of magnitude;
\item Compton equilibrium is not necessarily maintained at each point and the spectrum needs to be calculated in a self consistent way;
\item Klein- Nishina corrections to the Compton cross section must be included;
\item Relativistic corrections to the photon generation rate and spectrum (e.g. electron-electron Bremmstrahlung) need to be included.
\end{itemize}
The presence of high velocities require additional modifications:
\begin{itemize}
\item The radiation field is far from being isotropic and the diffusion approximation is not valid;
\item The hydrodynamic solution of the shock propagation as it approaches the surface requires a relativistic treatment.
\end{itemize}
The velocity effects are particularly severe if the shock reaches ultra-relativistic speeds.

The steady state structure of the relevant relativistic shocks has been solved numerically for ultra-relativistic shocks by \cite{Budnik10}, and its structure is approximately understood analytically \cite{Katz10,Budnik10,NakarSari12}. The time dependent problem of a relativistic breakout has been addressed using rough analytical arguments by \cite{NakarSari12}, who provided order of magnitude estimates for the emitted spectrum and light curve and its scaling properties with the breakout properties. As far as we know, an accurate calculation of a relativistic breakout, similar to the non-relativistic breakout calculations described in section \ref{sec:Planar}, is yet to be preformed. Given the complexity of the problem we believe that such a calculation is necessary for confirming the order of magnitude estimates of \cite{NakarSari12}, as well as for providing accurate predictions.

\section{Breakout and post breakout observations}
\label{sec:observations}

\subsection{Stellar surface breakouts and post breakout cooling emission}
\label{ssec:surface_obs}

\subsubsection{Stellar surface breakouts}
\label{sssec:surface_obs}

Shock breakout from a stellar surface is expected to produce a flash of X-rays with photon energies in the range $50-10000$~eV, and total energy, $E\approx 2\times 10^{47}R_{13}^2\vt_{\rm bo,9}\rm erg~s^{-1}$ (See Eq. \ref{eq:PlanarEinfty}) emitted over a time scale of tens of seconds to a few hours.  While the vast majority of supernovae are expected to have a non-relativistic shock breakout burst, a certain detection is yet to be found (see discussion below of the best candidate- the x-ray burst XRF080109 associated with the supernova SN2008D).

For large RSG progenitors, where the emitted energy is largest, the expected temperatures, of tens of eV, are such that most of the radiation is expected to be absorbed in the ISM. Smaller progenitors may produce $\gtrsim \rm keV$ emission with smaller outputs. Probably the most easily detectable breakouts are in supernovae with 'intermediate size', blue-super-giant (BSG) progenitors such as SN1987A \cite{Calzavara04,SKWspec13}. The current and past X-ray telescopes with highest potential detection rate of breakout bursts are ROSAT and XMM-Newton which have similar values of $A_{\rm eff}^{1.5} \times FOV\sim10^4\rm cm^{1.5}deg^2$, where $A_{\rm eff}$ is the effective area ($\approx 200\rm cm^2$ for ROSAT and $\approx 1000\rm cm^2$ for XMM-Newton) and $FOV$ is the field of view ($3.6\rm deg^2$ for ROSAT and $0.2-0.3 \rm deg^2$ for XMM-Newton, see comparison of different detectors in table 3 of \cite{SKWspec13}). XMM-Newton has much higher resolution and thus lower background making it the best detector so far for finding breakout bursts. Indeed, the expected number of background events within a single pixel, assuming a background X-ray flux of $n_{\rm background}0.01\rm counts~cm^{-2}~deg^{-2}~s^{-1}$ and an integration time of $1000\rm s$ is $0.1$ for XMM-Newton (PSF of $10^{-5}\rm deg^{-2}$) and $4$ for ROSAT (PSF of  and $0.002\rm deg^{-2}$). We are not aware of an attempt to systematically search for breakout bursts in the archival data of XMM-Newton even though detectable burst are expected toe exist in the data \cite{Calzavara04,SKWspec13}.

The total emitted energy in the breakout burst from a BSG is given by (see Eq. \ref{eq:PlanarEinfty})
\begin{equation}
\EBO=4\times 10^{46}\left(\frac{R}{3\times 10^{12}\rm cm}\right)^{2}\left(\frac{\vBO}{2\times 10^9 \rm cm~s^{-1}}\right)
\end{equation}
where the radius and breakout velocity are normalized to values expected to SN1987A,  $R\sim 3\times10^{12} \rm cm$, $\vBO\sim 2\times 10^9\rm cm~s^{-1}$ (see discussion in \cite{SapirHalbertal14} and references therein). The total number of emitted photons is thus expected to be
\begin{equation}
N_{\gamma}=\frac{\EBO}{\langle h\nu\rangle}=2\times 10^{55} \left(\frac{\EBO}{10^{46.5}\rm erg}\right)\left(\frac{\langle h\nu\rangle}{\rm 1 keV}\right)^{-1}.
\end{equation}

The expected number of photons is likely lower due to some absorption in the ISM.
Assuming $n_{\rm det}$ photons are required for a detection, the co-moving radial distance to which such an event can be seen using a detector with an effective area $A_{\rm eff}$ is
\begin{equation}
d=2900 \left(\frac{N_{\gamma}}{10^{55}}\right)^{1/2}\left(\frac{A_{\rm eff}}{1000\rm cm^2}\right)^{1/2}\left(\frac{n_{\rm det}}{10}\right)^{-1/2}\rm Mpc
\end{equation}
which corresponds to a redshift of order $z\sim 1$ (assuming $H_0=70\rm~km~s^{-1}~Mpc^{-1}$, $\Omega_M=0.3$ and $\Omega_{\Lambda}=0.7$, $z=1$ corresponds to  $d=3300$ Mpc).  Note that the number of photons arriving at the detector depends on the co-moving radial distance, unlike the instantaneous energy flux which depends on the luminosity distance. Note also that some suppression is expected given that the photons will be redshifted and some will not fall within the detector's band.

The rate of BSG SNe is  about a fraction of $f_{\rm BSG}\sim 0.01-0.03$ of 'Core Collapse' (CC) SNe \cite{Smartt09,Kleiser11,Pastorello12,Taddia16}. The rate of CC SNe at redshift of $z<1$ is estimated to be (see figure 10 of \cite{MadauDickinsonSFH14}, and references therein)
\begin{equation}
\dot n_{\rm CC}\sim (1+z)^3\times 10^{-4}\rm Mpc^{-3}~yr^{-1},
\end{equation}
The expected observable rate of BSG with $0.5<z<1$ and more than 10 photons is therefore
\begin{equation}\label{eq:XMMNewton}
\dot N_{\rm BSG}\sim 1.5\left(\frac{\dot n_{\rm BSG,eff}}{3\times 10^{-6}\rm Mpc^{-3}~yr^{-1}}\right)\left(\frac{A_{\rm eff}}{1000\rm cm^2}\right)^{3/2} \left(\frac{\rm FOV}{0.2\rm deg^2}\right)\left(\frac{N_{\gamma,\rm eff}}{10^{55}}\right)^{3
/2}\rm yr^{-1}.
\end{equation}

Given the $>10$ year lifetime of XMM-Newton, with an effective area of $\approx 1000 \rm cm^2$ at $h\nu=$keV and a field of view of $0.2\rm deg^2$ (exact values depending on the mode and frequency \cite{XMMHandbook}) several events may exist in the data. Evidently, this estimate has large uncertainties (BSG SNe rate and properties, absorption) and the number of events may range from none to tens. The result in \eqref{eq:XMMNewton} assumes a redshift of $0.5<z<1$ and ignores modest corrections within this range. We note that few attempts to identify $\sim100$ second time scale X-ray bursts in the archival data of ROSAT have been made with the aim of detecting Gamma-ray burst afterglows\cite{Vikhlinin98,Greiner00}. While most bursts are associated with M stars,  the possibility that a few of them are breakouts has not been ruled out to the best of our knowledge.

Future X-ray missions such as eRosita \cite{eRosita12},  HXMT \cite{Xie15HXMT} and Einstein Probe \cite{EinsteinProbe15} may significantly increase the prospects for detecting breakouts. In particular,  wide field detectors such as Einstein Probe with its $60{\rm deg}\times 60{\rm deg}$ field of view, may allow the detection of breakout bursts which can be later associated supernovae detected by deep wide field optical surveys such as the planned ZTF, and LSST.

Perhaps the best shock breakout candidate is the high energy X-ray flash (XRF080109) preceding the type Ib supernova (SN2008D) that was serendipitously discovered by the SWIFT  X-ray telescope during an observation of the NGC 2770 galaxy \cite{Soderberg08D}. The XRF had a total emitted energy of  $E \sim 2.5\times 10^{45}~\rm erg$ and duration of $\sim 300~\rm s$. The association with a supernovae and the fact such energies and time scales are within the range of shock breakouts lead several authors to suggest a shock breakout origin  (e.g. \cite{Soderberg08D,ChevalierFransson08,Katz10}). The fluence spectrum of the burst is hard and is consistent with a power law $\nu F_{\nu} \propto \nu^{0}$ \cite{Soderberg08D,Modjaz09}. The spectrum can also be fitted by the expected fluence spectrum of spherically symmetric breakouts \cite{SKWspec13}, with $h\nu_{\rm peak}\approx 4~\rm keV$, with some tension at the highest energy bins. The velocity inferred from this peak photon energy  implies $\vBO\sim 0.15 c$ and fastest parts of the ejecta moving at $\sim 0.3c$ which is consistent with radio observations  \cite{Soderberg08D}. Assuming a He envelope and $\kappa=0.2$, the progenitor's radius can then be found using the total emitted energy (see \eqref{eq:PlanarEinfty})
\begin{equation}
R\approx 5\times 10^{11}~\rm cm.
\end{equation}
The inferred radius is larger than the stellar radius $R_*\approx10^{11}$~cm inferred from the post breakout emission \cite{RW11} (see \S~\ref{sssec:post_bo_obs}), and from the radii of WR progenitors typically associated with this type Ib SN.
This suggests that the breakout may have taken place within an extended distribution of matter around the star \cite{Soderberg08D,Katz10,SvirskiNakar_08D_14} (a wind or a non-strandard outer envelope; recall that the mass lying ahead of the breakout radius is only $\sim 4\pi R^2 c/\vt \kappa\sim 10^{-7}M_\odot$, while the emission of post-breakout cooling radiation on a fraction of a day time scale is from shells of mass $\sim 10^{-3}M_\odot$).
The breakout explanation is challenged by the fact that the implied light-crossing time, $R/c$, is about 20 times smaller than the observed burst duration implying (within the breakout interpretation) either a non-spherical shock wave reaching different points on the surface at different times, or that the stellar envelope is enshrouded by a moderately optically thick circumstellar material (CSM) \cite{Soderberg08D,Katz10,SvirskiNakar_08D_14}. In particular, \cite{SvirskiNakar_08D_14} claim that a single model involving a dense wind can explain the time scale, energy scale and the observed spectrum using an approximate analytic calculation \cite{SvirskiNakar14}.

The optical, low energy tail of shock breakout may be detected with sufficient cadence and sensitivity. Two type II-P supernovae were recently reported to be discovered in the data of the planet transit search mission KEPLER, with 30-minute cadence and excellent photometric accuracy (\cite{Garnavich16}, the SN identification is based on light curves alone). In particular, the early rise of the light curve of one of these supernovae (KSN2011d) seems to be consistent with having a breakout burst lasting a few hours. While the data does not allow a clear conclusion (in our view), this work demonstrates that the required sensitivity has been roughly reached, motivating future similar searches.

\subsubsection{Post breakout cooling emission}
\label{sssec:post_bo_obs}

Following breakout, the expanding cooling envelope produces a bright, $L\sim10^{43}{\rm erg/s}$, UV emission on a time scale of hours to days (\S~\ref{sec:spherical}). As discussed in some detail in \S~\ref{sec:ObsTheory}, a measurement of the color temperature enables one to determine $R_*$, and a measurement of $L$ enables one to determine $R_*E/M$. The simple model described in \S~\ref{sec:spherical} applies up to $\sim1(M/M_\odot)^{1/2}$~d (assuming $E_{51}/(M/M_\odot)\simeq1$, see eq.~(\ref{eq:t_valid})). Since the temperature at this time is $\sim1$~eV (see eqs.~(\ref{eq:non_rel_T_ph}),~(\ref{eq:EffModel-THe}),~(\ref{eq:EffModel-TCO})), corresponding to a $\lambda f_\lambda$ peak at $\sim0.3\mu$, this implies that UV observations at early time are required in order to determine $R_*$.

Early, $\le1$~d, observations of SN lightcurves became available recently with the beginning of the operation of wide field sensitive optical surveys, like the Palomar Transient Factory (PTF, \cite{Rau_PTF09,Law_ZPTF09}, and iPTF, \cite{Gal-Yam_iPTF11}), the Panoramic Survey Telescope and Rapid Response System (Pan-STARRS, \cite{Pan-STARRS02}) and the All-Sky Automated Survey for Supernovae (ASAS-SN; Shappee et al. 2014). The sensitivity of these surveys allows one to detect post-breakout cooling emission on a day time scale from large, RSG, progenitors, and to set upper limits on the emission from smaller, BSG/WR progenitors (recall that $L\propto ER_*/M$). In most cases, early UV observations (from space) are not available, hence limiting the ability to constrain $R_*$. The rate of detection of SNe at $t\le$1~d will increase as new surveys become operative, like SkyMapper \cite{SkyMapper07}, the Zwicky Transient Facility (ZTF, \cite{Law_ZPTF09}), and the Large Synoptic Survey Telescope (LSST, \cite{LSST09}). ZTF will provide, for example, 10/2/1 RSG/BSG/WR shock cooling detections per year at $t<1$~d at g-band, and LSST will roughly double this rate \cite{Ganot14}. This will significantly improve the ability to constrain models. However, a real breakthrough would require wide field space UV observatory like the proposed ULTRASAT satellite \cite{Sagiv14}, which will provide a tenfold increase in detection rate compared to the ZTF \cite{Ganot14} and, most crucially, will provide early UV measurements.

The main constraints inferred from observations of post-breakout shock cooling emission are briefly summarized below.
\begin{itemize}
  \item \emph{Type II SNe}.
    \begin{itemize}
    \item A simultaneous PTF (optical) and GALEX (space, NUV) search for early UV SNe emission resulted in the detection of 7 SNe of type II, typically at $\sim3$~d past the explosion \cite{Ganot14}. The observations are consistent with explosions of RSGs, with $R_*\sim3\times10^{13}$~cm and $E/M\sim0.1\times10^{51}/M_\odot$. However, the quality of the data is not sufficient for accurately inferring $R_*$ and $E/M$. Three earlier examples of UV emission on $\sim2$~d time scale from type II SNe, two serendipitous detections by GALEX \cite{Gezari08,Schawinski08} and one resulting from a coordinated GALEX-Pan-STARRS search \cite{Gezari10}, yielded similar conclusions. Note that in earlier papers a clear distinction between "shock breakout" emission and "post breakout cooling" emission was not made. Hence, although the UV emission is referred to in refs.~\cite{Gezari08,Schawinski08} as "shock breakout" emission, it is probably related to the post-breakout cooling phase \cite{RW11,Gezari10}.
    \item A PTF search for early SN emission yielded detections of 57 SNe of type II, with good R-band sampling at $t <10$~d \cite{Rubin15}. These observations lead to determinations of $E/M$ to within a factor of 5, with an average of $0.1\times10^{51}{\rm erg}/M_\odot$ and a positive correlation of $E/M$ with $^{56}$Ni mass, and yielded only weak constraints on $R_*$ (note that in the RJ regime, $\nu L_\nu\propto T^{-3}L\propto(E/M)R^{-1/4}$).

        Comparing a large sample of SNII lightcurves to the post-breakout cooling model predictions, radii much smaller than expected for RSG progenitors were inferred in refs.~\cite{Gall15,Gonzalez15}. However, this conclusion is obtained by comparing the data to the model well beyond the model's validity time \cite{Rubin15}. Comparing multi-band light curves of two individual SNII to the model prediction of \cite{RW11}, but limiting the analysis to $t<1$~week, radii consistent with $\sim10^{13.5}$~cm are inferred in refs.~\cite{Valenti14,Bose15}.
    \end{itemize}
  \item \emph{Type Ib/c SNe}.
    \begin{itemize}
      \item The non-detection of post-breakout cooling emission in observations of two SNe of type Ic (PTF 10vgv, 1994I) \cite{Corsi_WR12} and one SN of type Ib (iPTF13bvn) \cite{Cao_WR13} was used to set upper limits on the progenitor radii of $R_*/R_\odot<(1,0.25,{\rm few})$ respectively, implying WR progenitors (or CO cores of stars stripped in binary systems).
      \item There are two examples in which a serendipitous detection by the SWIFT satellite of an X-ray/$\gamma$-ray flush preceding the UV/optical emission of a type ${\rm Ib/c}$ SN lead to early space UV/O observations of the SN emission: the low-luminosity GRB (LLGRB) GRB 060218 ($E_\gamma\sim10^{49}$~erg) associated with SN2006aj \cite{Campana06,Pian06aj,Mazzali06aj,Soderberg06aj}, and the X-ray flash (XRF) XRO080109 ($E_X\sim10^{46}$~erg) associated with SN2008D \cite{Soderberg08D,Mazzali08Sci}. The preceding LLGRB/XRF have been suggested to be generated by a breakout through a wind, and will thus be discussed in the following sub-section. The UV/O emission observed on a $\sim1$~d time scale is consistent with post-breakout cooling emission \cite{Campana06,WaxmanCampana07,Soderberg08D,ChevalierFransson08,RW11}. As explained at the end of \S~\ref{sec:ObsTheory}, the early emission of SN2008D was used to determine the progenitor's radius, $R_*\approx10^{11}$~cm, $E/M$, $E_{51}/(M/M_\odot)\approx0.8$, and the reddening, $E(B-V)=0.6$ \cite{RW11}. A detailed analysis of this type was not carried out for SN2006aj.
    \end{itemize}
  \item \emph{Type Ia SNe}. A recent review of the observational constraints on the progenitors of Ia SNe may be found in ref.~\cite{Maoz_Ia_14}. Here we briefly describe the main aspects related to very early observations.
    \begin{itemize}
    \item The non-detection of post breakout cooling emission has been used to put stringent constraints on the radii of the progenitors, of order $0.1R_\odot$ \cite{Nugent11fe,Bloom11fe,Zheng13Ia,Im15Ia}, strongly constraining the possible progenitors. The main factor contributing to the uncertainty in this limit is the uncertainty in the determination of the explosion time. The most stringent limit, $R_*/R_\odot<0.05$ was obtained for SN2011fe \cite{Nugent11fe,Bloom11fe,PiroNakar_Ia13,PiroNakar_Ia14,Mazzali14fe}.
    \item The collision of the expanding SN ejecta with a stellar companion may lead to significant emission of radiation, and hence to deviations from a "standard" early light curve. Additional deviations may be due to mixing of Ni in the outer envelope. These topics are beyond the scope of this review, and we refer the reader to ref.~\cite{Maoz_Ia_14} for a detailed discussion.
    \end{itemize}
  \item \emph{"Double-peak SNe"}. The bolometric light curves of several SNe, mainly of the IIb class \cite{Wheeler93J,Arcavi11dh,VanDyk14df} (super-luminous double-peaked SNe are discussed in the senc subsection), show a "double peak" behavior: a first peak at a few days after the explosion, preceding the main SN peak (on time scale of tens of days). It is commonly accepted that the first peak is produced by the post-breakout shock cooling radiation from an extended, $R_*\sim10^{13}$~cm, low mass, $M\le0.1M_\odot$ envelope \cite{Woosley9493J,Bersten12dh,NakarPiro14,Piro15}. Such a low mass shell would become transparent after a few days of expansion, producing a first peak in the light curve well before the time at which the bulk of the ejecta becomes transparent. The model described in \S~\ref{sec:spherical} does not apply, of course, up to times at which the envelope becomes transparent, see e.g. eq.~(\ref{eq:t_valid}), and cannot therefore describe the behavior near the bolometric peak. However, it should apply to the early rising part of the lightcurve, since this part is not sensitive to the details of the density structure.
\end{itemize}

\subsection{Extended CSM breakouts}
\label{ssec:CSM_obs}

Breakouts from an extended CSM at large radii are very bright on days time scale (see eq.~(\ref{eq:CSM_scales})). The main observational challenge to inferring stringent constraints on the progenitors and on their environment is the lack of UV/X-ray measurements, which are required in order to determine the characteristic plasma parameters. As explained in some detail in \S~\ref{ssec:CSM}, a major additional challenge is the lack of a complete quantitative model describing the spectra of the emitted radiation. In particular, it is difficult to determine the density distribution and the origin of the extended CSM (winds, pre-ejected shells), and also whether or not late injection of energy into the expanding ejecta plays a significant role (see below). These gaps are reflected in the following discussion of observations of CSM breakouts.

\begin{itemize}
  \item \emph{Type IIn/Ibn SNe}. There is significant and increasing evidence that the early lightcurves of SNe of these types are generated by wind breakouts \cite{Ofek14IIn-wind,Drout14Wind,Gezari15Wind}. However, the inferred mass loss, $\sim10^{-3}M_\odot/{\rm yr}$, is typically higher than expected in stellar evolution models \cite{Ofek14IIn-wind,Ofek14extremeMdot}, and there are discrepancies between the observed and the predicted X-ray emission \cite{Ofek13WindX}. The lack of complete self-consistent theoretical models does not allow ruling out other models and hinders the inference of stringent quantitative constraints (quoting ref.~\cite{Ofek13WindX}: "We still do not have a good theoretical understanding of the expected X-ray spectral evolution... our observations cannot yet be used to rule out other alternatives"). These conclusions are based on the following main observations.
    \begin{itemize}
      \item The lightcurves of 15 IIn SNe observed by PTF on times scales of $\sim10$~d are consistent with CSM breakouts with $\vt_{\rm bo}\sim 10^9 {\rm cm/s}$ and mass loss rates of $\dot{M}\sim10^{-3}M_\odot/{\rm yr}$ \cite{Ofek14IIn-wind} (analyzing the same data, it is concluded in ref.~\cite{Moriya14Mdot} that the $\dot{M}$ distribution is wide, spanning an order of magnitude).
      \item Observations of 12 PAN-STARRS transients \cite{Drout14Wind}, with characteristic $L\sim10^{43}{\rm erg/s}$ and rise times $<10$~d, are consistent with wind breakouts with $R_{\rm bo}\sim3\times10^{14}$~cm and $T>1$~eV (or with breakouts from “non-standard” extended low mass envelopes with similar breakout parameters).
      \item X-ray (XRT and Chandra) observations of 19 SNe of type IIn and one of type Ibn \cite{Ofek13WindX} yielded mixed conclusions regarding the wind breakout origin of these events, as some were consistent and some too bright for CSM breakouts.
    \end{itemize}
  \item \emph{Super luminous (SL) SNe}.
    \begin{itemize}
      \item A few dozens examples are known (see \cite{Gal-Yam12SLSN} for a review) of extremely bright, $L>10^{44}{\rm erg/s}$, SNe of type II (H rich) and I(c) (H poor) (we do not discuss here SLSN of type R, which are likely powered by radioactive decay). The observed radiation may be interpreted as a breakout from extended CSM, with $R_{\rm bo}\sim10^{15}$~cm and $\vt_{\rm bo}\sim10^9{\rm cm/s}$, which is H-rich/poor for SLSN-II/I(c) (e.g. \cite{OfekWind10,Balberg11,GinzburgBalberg12,Moriya13SL}). The origin and structure of the CSM (extended envelope, wind, pre-ejected shell) are not well constrained. An alternative type of models was suggested, in which the expanding SN ejecta is continuously heated as it expands by a long lasting "central engine" such as a "magnetar" or an accreting black hole \cite{Maeda07magnetar,KasenBildsten10Magnetar,Woosley10Magnetar,Dessart12magnetar,Dexter13BH,Inserra13Magnetar,Nicholl13magnetar,Howell13magnetar,Metzger15magnetar}. In these models, the deposition of thermal energy in the ejecta at large radii circumvents the adiabatic losses due to the large expansion factors. Finally, we note that "quark-nova" modes have also been suggested as an explanation of SLSNe \cite{Leahy08QN,Ouyed12QN}.
      \item Several SLSN of type I with double-peak bolometric lightcurves have been recently reported \cite{Leloudas1206oz,Nicholl15bdq,Smith15DES14X3taz,Nicholl15dpSL}. Similarly to the type IIb double-peaked events (see preceding sub-section), the first peak is commonly interpreted as the post-breakout cooling emission from an extended CSM, although it is not clear whether or not this material is part of an extended stellar envelope (e.g. \cite{Piro15}). The origin of the second peak is debated, with tendency to prefer models with a "central engine" heating as the second peak driver. Proponents of the central engine magnetar models have furthermore suggested that the first peak may due to shock breakout from ejecta that was inflated to large radius by the energy output of the magnetar \cite{Kasen15SLdp}.
    \end{itemize}

   \item \emph{Low luminosity GRBs (LLGRBs) and X-ray flashes (XRFs)}.
    \begin{itemize}
      \item It has been suggested, based mainly on qualitative order of magnitude analyses, that LLGRBs and XRFs associated with SNe are produced by shock breakouts \cite{Kulkarni98bw,Tan01,Campana06,Li07,WaxmanCampana07,Katz10,NakarSari12,Nakar15aj}, possibly through extended CSM environments. The high temperatures of the bursts (tens - hundereds of keV, see \eqref{eq:PlanarVTanalytic}) and properties of the later radio and X-ray emission suggest that if true, these breakouts require relativistic corrections making quantitative estimates difficult.  A rough analytic estimate of the properties of relativistic breakouts was carried out in \cite{NakarSari12} based on the properties of relativistic radiation mediated shocks \cite{Budnik10,Katz10}, leading to relativistic corrections to the breakout temperatures, energies and durations. With these corrections, the long duration $t\sim 1000\rm s$  LLGRBs associated with SNe, GRBs 060218- SN2006aj \cite{Campana06} and GRB 100316D-SN2010bh \cite{Starling11} as well as the short duration $t\sim 30\rm s$ LLGRBs associated with SNe, GRB 980425-SN 1998bw \cite{Galama98},   GRB 031203- SN 2003lw \cite{Malesani04} were shown to be broadly consistent with spherical breakouts from a spherical surface. The radius required by the longer LLGRBs is quite extended $\gtrsim 10^{13}$ suggesting a CSM origin (e.g. \cite{Campana06}) or an extended progenitor (\cite{Nakar15aj}, allowing a possible unification model with cosmological GRBs). Alternatively, the long duration may be a result of a significant departure from spherical symmetry (e.g. \cite{WaxmanCampana07}).
      \item Due to the uncertainties in the model, and to the fact that existing analyses do not usually account for all of the observed radiation components, there is no consensus regarding this interpretation and various alternative models are being discussed, mostly involving the presence of relativistic jets (see e.g. a recent discussion in \cite{IrwinChevalier15aj}).
    \end{itemize}
\end{itemize}

\bibliographystyle{spphys}

\end{document}